\documentclass{docclass} 

\allowdisplaybreaks

\lefttitle{F. Aerts, D. Nuyens and J. Meyers}
\righttitle{A probability distribution in the phase space of turbulent channel flow} 

\title{A machine-learned probability distribution in the phase space of turbulent channel flow for synthetic turbulence and flow reconstruction}

\author{Frederik Aerts\aff{1}, Dirk Nuyens\aff{2} \and Johan Meyers\aff{1}}

\affiliation{\aff{1}Department of Mechanical Engineering, KU Leuven, Leuven, Belgium
\aff{2}Department of Computer Science, KU Leuven, Leuven, Belgium}

\corresau{Frederik Aerts, frederik.aerts1@kuleuven.be}

\begin{document}
\maketitle

\begin{abstract}

 \noindent
Although a complete characterisation of the probability distribution in the phase space of turbulent flows remains elusive, accurately sampling this distribution is crucial for both synthetic turbulence generation and turbulent flow reconstruction. 
Motivated by these applications, we examine to what extent a machine-learned distribution can approximate the physical invariant distribution of turbulent channel flow at $\Rey_\tau=180$. 
We evaluate the approximation in terms of three properties: the reproduction of essential turbulent statistics, the consistency of conditional sampling, and its dynamical invariance. 
To this end, a flow-based generative model is trained on a minimal conditional flow unit, which we define as the smallest domain outside which conditional fields, given a single observation at the domain centre, are indistinguishable from unconditional fields in terms of mean-square discrepancy to other conditional fields. 
We also introduce an approximate procedure for conditional sampling from the learned distribution. 
Comparisons with direct numerical simulation show that synthetic turbulent fields reproduce key statistical and dynamical features of turbulence, including scale-dependent intermittency and interscale energy transport. 
The consistency of conditional sampling is demonstrated in a flow reconstruction problem, and subsequently used to generate synthetic turbulent velocity fields on a large domain. 
When adopted as initial conditions in direct numerical simulations, these fields yield physical and statistically stationary ensemble statistics, indicating that the learned distribution provides a good approximation to the natural distribution of the turbulent dynamical system.

\end{abstract}


\section{Introduction \label{sec:intro}}

The probability distribution in the phase space of the velocity field is central to the statistical description of incompressible turbulence \citep{monin_statistical_1971}. 
Its ensemble statistics, such as the Reynolds stresses, energy spectra, and structure functions, describe the structure of turbulence and have been widely studied in theory, experiments, and simulations \citep{monin_statistical_1971, frisch_turbulence_1995, pope_turbulent_2000}. 
Within the dynamical-systems perspective on turbulence pioneered by \cite{hopf_mathematical_1948}, it is conjectured that this probability distribution should remain invariant under the turbulent dynamics \citep{hopf_statistical_1952}. 
To be physically relevant, it should furthermore be selected as the push-forward of typical initial conditions \citep{ruelle_what_1978,eckmann_ergodic_1985}.
The resulting physical invariant distribution is known as the natural distribution of the dynamical system \citep{cvitanovic_chaos_2020}.
If this natural distribution formally exists, it is supported on the attractor on which the dynamics live, 
which is typically fractal and thin in phase space for dissipative chaotic systems such as turbulence \citep{eckmann_ergodic_1985}. 
The complete characterisation of the presumably highly intricate natural distribution of any turbulent flow remains elusive, reflecting the outstanding `problem of turbulence' \citep{monin_statistical_1971}.  
Fortunately, a complete determination of the natural distribution is by no means necessary in most applications. However, applications such as synthetic turbulence generation and turbulent flow reconstruction inherently rely on our ability to sample this natural distribution.

The premise of synthetic turbulence generators is to construct velocity fields that reproduce turbulent ensemble statistics, without simulating the underlying dynamics.
When adopted as initial or inflow conditions for numerical simulations of turbulent flows, the synthetic fields should reduce the computational cost required to reach statistical stationarity. 
In this respect, the statistics of an ensemble of synthetic flow fields should precisely satisfy the dynamical invariance property of the natural distribution. Classical synthetic random Fourier methods build on the construction of divergence-free Gaussian random fields with a prescribed energy spectrum by \cite{kraichnan_diffusion_1970}. The anisotropy and inhomogeneity of wall-bounded turbulence with mean shear can be incorporated through rapid distortion theory \citep{mann_spatial_1994}, or by transforming the field with the Cholesky decomposition of the Reynolds stress tensor \citep{lund_generation_1998} and rescaling it \citep{smirnov_random_2001}.
Alternatively, one can start from white noise in real space and filter it with the spatial correlation function, or an approximation of it, to impose an energy spectrum on the turbulent fluctuations \citep{klein_digital_2003,di_mare_synthetic_2006,xie_efficient_2008}. 
Although such approaches yield the correct energy content at all scales, they do not necessarily capture the interscale energy transfer. The latter depends on third-order moments of the velocity increments \citep{vassilicos_dissipation_2015}, which are exactly zero for Gaussian fields. 
In addition, Gaussian fields lack the statistical predominance of vortex stretching and self-amplification of the strain rate, as well as scale-dependent intermittency, necessitating an adaptation period in the simulation during which a realistic turbulent field develops \citep{rosales_minimal_2006}. 
Other notable routes to generate synthetic turbulent fields are the application of a minimal multiscale Lagrangian map to an initial Gaussian field  \citep{rosales_minimal_2006}, and the superposition of coherent structures in synthetic-eddy-methods \citep{jarrin_synthetic-eddy-method_2006, poletto_new_2013}. However, such approaches do not generally provide a tractable conditional-sampling procedure for their induced distributions.

Turbulent flow reconstruction methods aim to efficiently find turbulent velocity fields that are consistent with typically sparse observations and to quantify the uncertainty on these reconstructions.
In the absence of measurement noise, the problem of instantaneous flow field reconstruction is equivalent to sampling the natural distribution conditioned on the observed variables. 
One of the first approaches to address this challenge was Linear Stochastic Estimation (LSE) \citep{adrian_role_1977, adrian_stochastic_1988}. LSE reconstructs the best linear estimator for the conditional average of the flow field given the observations, with a procedure that is numerically equivalent to approximating the conditional natural distribution with a Gaussian \citep{adrian_stochastic_1996}.
The conditional average of the flow field is optimal in terms of mean-square reconstruction error.
However, the resulting so-called `conditional eddies' do not necessarily capture the instantaneous structures present in the actual flow field: they represent simply the average field associated with the observed event \citep{adrian_stochastic_1996}. 
LSE and its spectral variant \citep{tinney_spectral_2006} are not only widely used to retrieve such conditional structures, but also to address fundamental questions, such as the observability of turbulent structures from wall measurements for instance \citep{encinar_logarithmic-layer_2019}. 
Another classical approach, known as gappy Proper Orthogonal Decomposition (POD) \citep{everson_karhunenloeve_1995,venturi_gappy_2004, bui-thanh_aerodynamic_2004}, seeks reduced order representations of the reconstructed instantaneous fields, by only considering the dominant modes of the covariance tensor \citep{sirovich_chaotic_1989}. 
Similar to LSE, which can also be formulated in a POD basis \citep{bonnet_stochastic_1994}, gappy POD minimises the mean-square reconstruction error, but with a different regularisation of the POD coefficients. 
Other recently employed regularisation methods are sparsity promoting regularisation \citep{callaham_robust_2019} and Tikhonov regularisation \citep{bertram_fusing_2021}.
Although these approaches can achieve low reconstruction errors, they generally do not reconstruct realistic instantaneous fields, in the sense that the statistics of the reconstructed flow field deviate from the true ensemble statistics.

\begin{figure}
    \centering
    \includegraphics[width=0.9\linewidth]{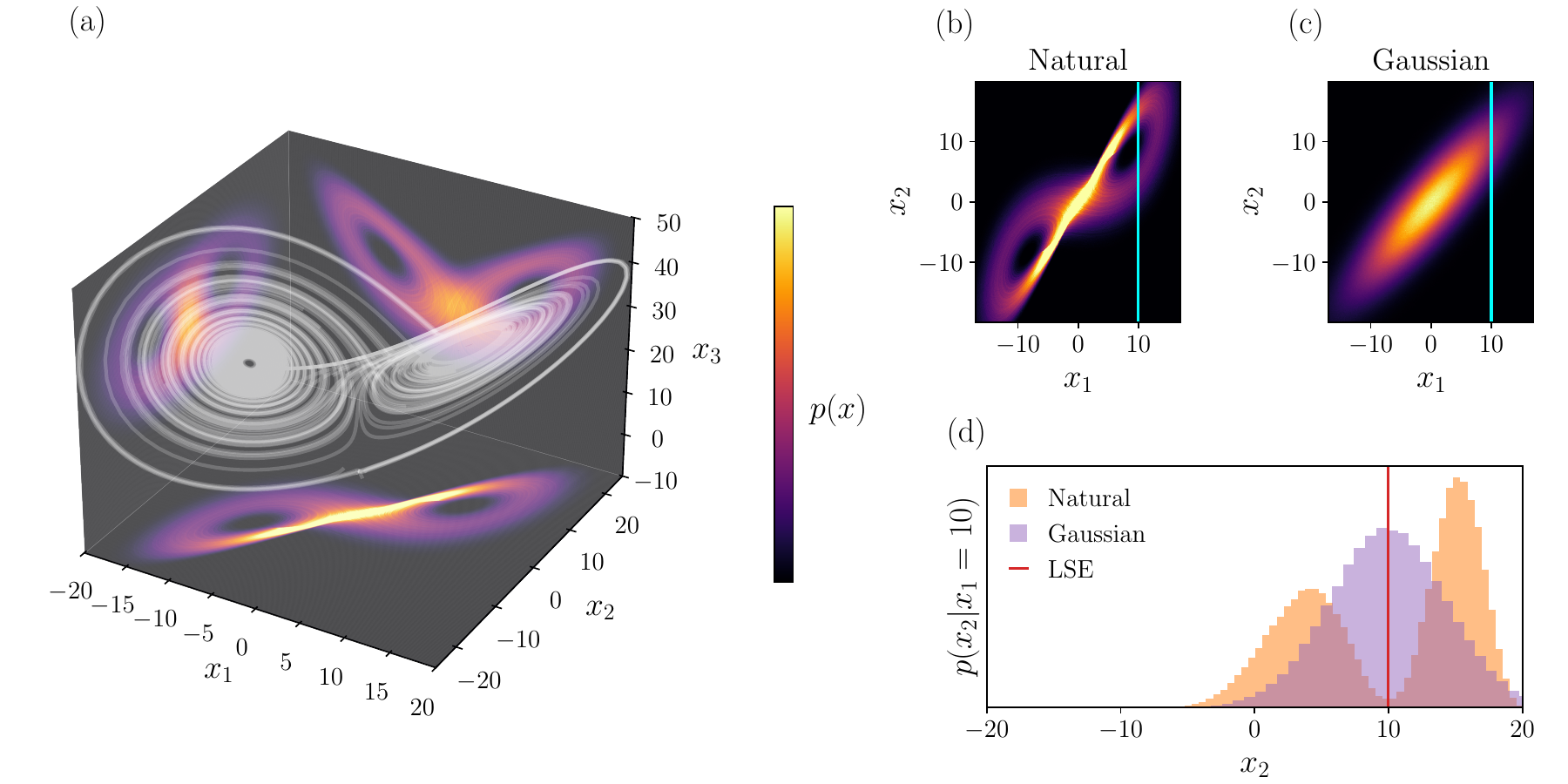}
    \caption{Illustration of the natural distribution for the dynamical system representing cellular convection by \cite{lorenz_deterministic_1963}: {(a)} invariant distribution obtained by the push forward of typical initial conditions, {(b)} marginal distribution $p(x_1, x_2)$ derived from the natural distribution, (c) Gaussian approximation to this marginal distribution, (d) reconstruction of $x_2$ given $x_1=10$ based on samples of the natural distribution, a Gaussian approximation to it, and Linear Stochastic Estimation (LSE), which equals the conditional mean of the Gaussian. }
    \label{fig:lorenz}
\end{figure}

The aforementioned synthetic turbulence generators and reconstruction methods focus mostly on the first- and second-order statistics of the natural distribution. 
That such approaches can yield a relatively poor approximation to the natural distribution of a chaotic system may be demonstrated with the prototypical dynamical system by \cite{lorenz_deterministic_1963}.
This dynamical system represents a simplified form of cellular convection, in which the velocity field is described by a mode with amplitude $x_1$ and the temperature field by two modes with amplitudes $x_2$ and $x_3$. Their dynamics are given by
\begin{equation}
    \frac{\mathrm{d} }{\mathrm{d}t } \begin{pmatrix}
        x_1 \\ x_2 \\ x_3
    \end{pmatrix}  = \begin{pmatrix}
        \sigma (x_2 - x_1) \\ x_1 (\rho-x_3) - x_2 \\ x_1 x_2 - \beta x_3
    \end{pmatrix},
\end{equation}
where $ \sigma=10$ is the Prandtl number, $\rho=28$ is the Rayleigh number normalised by its critical value, and $\beta=8/3$ is a number related to the aspect ratio of the convective rolls. 
For a supercritical Rayleigh number $\rho>1$, the fixed point representing the absence of convection at $(0,0,0)$ becomes a saddle point and two other fixed points emerge at $(\pm \sqrt{\beta(\rho-1)}, \pm \sqrt{\beta(\rho-1)}, \rho-1)$. Those represent steady left- or right-turning convective rolls and also become saddle points for $\rho > \sigma (\sigma +\beta+3)(\sigma - \beta -1)^{-1}$ \citep{lorenz_deterministic_1963}. 
Although invariant distributions can be constructed as point masses at these fixed points, they are of little physical value, since they do not reproduce the ensemble statistics of typically observed trajectories. 
When starting from more `typical' points sampled from any other part in phase space, the trajectories evolve towards a strange attractor \citep{ruelle_nature_1971}.
Figure~\ref{fig:lorenz}(a) demonstrates how the natural distribution is obtained from the push-forward of initial conditions drawn from a multivariate Gaussian with an isotropic variance of 0.5 centred around the fixed point at the origin.
It is seen that the natural distribution for this system is distinctly non-Gaussian, as it is neither globally nor locally dense.
Suppose now that an observation of the velocity field is available, such that it is known that $x_1=10$, and let us be interested in reconstructing the temperature mode amplitude $x_2$.
Figures~\ref{fig:lorenz}(b) and (c) show the marginal natural distribution $p(x_1,x_2)$ and a Gaussian distribution with the same mean and covariance.
The conditional natural distribution $p(x_2 |x_1=10)$, shown in figure~\ref{fig:lorenz}(d), is obtained by retaining samples from the joint distribution in figure~\ref{fig:lorenz}(b) with $9.9 \leq x_1 \leq 10.1$. 
It shows that the dynamics dictate $x_2$ to be either somewhat larger or smaller than $10$.
However, this information is lost in the Gaussian approximation to the conditional distribution, which is obtained analogously from the joint distribution in figure~\ref{fig:lorenz}(c). 
Although its conditional mean, which corresponds to the reconstruction from LSE, yields a low reconstruction error, it is unlikely to be observed under the natural distribution. 
To synthesise and reconstruct realistic states, one should therefore sample an approximation of the natural distribution that captures higher-order statistics as well.

Recent advances in probabilistic machine learning have shown that generative models offer promising computational methods to generate and reconstruct instantaneous turbulent flow fields. 
Generative Adversarial Networks (GANs) let a generator neural network and a discriminator neural network compete in a zero-sum game to generate flow fields that cannot be discriminated from reference flow fields \citep{goodfellow_generative_2014,goodfellow_generative_2020}. These have for instance been used for super-resolution reconstruction \citep{xie_tempogan_2018, deng_super-resolution_2019,kim_unsupervised_2021,yousif_high-fidelity_2021}, synthetic turbulence generation of two-dimensional cuts of the velocity field \citep{kim_deep_2020,drygala_generative_2022} and complete fields of homogeneous isotropic turbulence \citep{mohan_gan_2020}, 
reconstruction of two-dimensional cuts of the velocity field from plane observations \citep{li_multi-scale_2023-1} and wall measurements \citep{guemes_coarse_2021},
and the reconstruction of spatially three-dimensional flow fields from plane observations \citep{yousif_deep-learning_2023} and wall measurements \citep{cuellar_three-dimensional_2024}. 
Diffusion models corrupt reference flow fields by iteratively adding noise and learning how to reverse this process \citep{sohl-dickstein_deep_2015,ho_denoising_2020,song_score-based_2021}, and have been shown to outperform GANs on image generation benchmarks \citep{dhariwal_diffusion_2021}.
Such models have for example been used for super-resolution reconstruction \citep{shu_physics-informed_2023}, the reconstruction of two-dimensional slices of turbulent flow fields \citep{li_multi-scale_2023,vishwasrao_diff-sport_2025}, and the generation of spatially three-dimensional turbulence \citep{lienen_zero_2023}. 
They have also been used for the generation and reconstruction of spatially three-dimensional turbulence from observational setups prescribed during training \citep{rybchuk_ensemble_2023} or imposed
during sampling \citep{amoros-trepat_guiding_2026}, as well as for the generation and super-resolution reconstruction of two-dimensional cuts \citep{gao_bayesian_2024} and for the generation and reconstruction of spatially three-dimensional spatio-temporal fields \citep{du_conditional_2024,steinbrenner_turbulence_2026}.
When comparing GANs and diffusion models for flow reconstruction in rotating turbulence, \cite{li_multi-scale_2023} showed not only that diffusion models outperform GANs, but also that there exists a trade-off between obtaining a reliable mean-square error and realistic ensemble statistics of the reconstructed part of the flow field. 
This motivates a probabilistic interpretation of the mean-square error, enabling the derivation of an expected error for instantaneous reconstructions.

Another class of generative models, known as flow-based generative models, learns a deterministic transformation that transports samples from a source distribution to a target distribution \citep{papamakarios_normalizing_2021}. This transformation may be viewed as a dynamical system in fictive time \citep{chen_neural_2019}, in which a distribution of initial conditions evolves to the target distribution through a prescribed sequence of intermediate distributions, termed a probability path. Individual samples follow trajectories generated by a phase-space vector field, analogous to fluid elements advected by a time-dependent velocity field. To distinguish this vector field from the physical velocity field, we refer to it as the generator. Flow matching provides an efficient framework for learning such generators from prescribed conditional probability paths, such as the conditional optimal transport path \citep{lipman_flow_2023,lipman_flow_2024}. When the prescribed probability path corresponds to that of a score-based diffusion model, flow matching learns the generator of the associated probability-flow ordinary differential equation \citep{song_score-based_2021}. More generally, variational diffusion \citep{sohl-dickstein_deep_2015,ho_denoising_2020}, score-based \citep{song_score-based_2021} and Gaussian flow-matching models \citep{lipman_flow_2023} are closely related in their parameterisations, probability paths and training objectives, but retain distinct theoretical and practical properties \citep{kingma_understanding_2023,lipman_flow_2024,lai_principles_2025}.
In particular, learning the generator with a conditional optimal-transport path offers favourable properties for both training and sampling \citep{lai_principles_2025}, and has been reported to improve training efficiency, sampling speed and generative performance relative to diffusion paths \citep{lipman_flow_2023}. A flow-matching model based on this path has recently also been applied to reconstruct velocity fluctuations at different wall-normal planes from wall measurements in turbulent channel flow \citep{parikh_conditional_2026}.

For probabilistic machine-learning models to approximate the natural distribution of a turbulent dynamical system, they should generate and reconstruct ensembles of three-dimensional turbulent fields that reproduce essential turbulent statistics and remain statistically invariant under the governing dynamics.
Assessments in homogeneous isotropic turbulence went beyond energy spectra to include higher-order statistics, such as the distribution of the $Q$ and $R$ invariants of the velocity-gradient tensor \citep{mohan_gan_2020} and the Q-criterion \citep{amoros-trepat_guiding_2026}. 
In homogeneous anisotropic rotating turbulence, scale-dependent flatness and vorticity distributions have likewise been examined in addition to spectra \citep{li_3d_2026}.
Evaluations in inhomogeneous and anisotropic wall-bounded turbulence have focused primarily on statistics up to second order \citep{du_conditional_2024,holzschuh_p3d_2025} and, more recently, on third- and fourth-order one-point statistics as well as distributions of the velocity increments and $Q$-$R$ statistics \citep{steinbrenner_turbulence_2026}. Nevertheless, key aspects of the energy dynamics responsible for maintaining the statistical state of wall-bounded turbulence, including the distinctly non-Gaussian interscale energy transfer, have not yet been assessed.
Despite being a defining property of the natural distribution, the dynamical invariance of such approximations has to the best of our knowledge not yet been investigated for a turbulent dynamical system.
A direct test is to initialise direct numerical simulations with an ensemble of synthetic fields and examine the subsequent evolution of their ensemble statistics. Such a test requires the synthetic fields to resolve all dynamically relevant scales and is preferably performed in a domain sufficiently large in the homogeneous directions. 
However, it largely remains an open question how generative models can be scaled to spatial domains much larger than those used in training \citep{wang_generative_2026}.

The objective of this work is to evaluate to what extent a machine-learned distribution can approximate the physical invariant distribution of a turbulent channel flow at a friction Reynolds number of $\Rey_\tau=180$.
We assess the approximation in terms of three properties relevant to synthetic turbulence generation and turbulent flow reconstruction: the reproduction of essential turbulent statistics, the consistency of conditional sampling, and its dynamical invariance.
To this end, a flow-based generative model is trained on what we call a minimal conditional flow unit, so that it can approximately represent all dynamically relevant scales on arbitrarily large domains. 
In addition, we introduce a new approximate technique to sample the learned conditional distribution.
When validating the consistency of conditional sampling, the mean-square reconstruction error is compared to an LSE-based estimated reconstruction error, obtained from its probabilistic interpretation.
To the best of our knowledge, this is the first attempt to evaluate a probabilistic machine-learning approximation of the physical invariant distribution of large-domain, inhomogeneous, anisotropic three-dimensional turbulence through the explicit assessment of both dynamical invariance and key turbulent statistics, including interscale energy transfer and scale-dependent intermittency.

The manuscript is further organised as follows. 
The machine-learned probability distribution is described in~§\ref{sec:methods}, covering the approximate representation of the natural distribution on a minimal conditional flow unit, the flow-based generative model, the new conditional sampling method, and the case set-up. 
In~§\ref{sec:results}, the three properties, namely turbulent ensemble statistics, consistent conditional sampling, and dynamical invariance, are examined through applications in synthetic turbulence generation and turbulent flow reconstruction. 
Conclusions are then drawn in~§\ref{sec:conclusion}.

\section{Machine-learned probability distribution}\label{sec:methods}
We consider a direct numerical simulation of incompressible pressure-driven channel flow at friction Reynolds number $\Rey_{\tau}=180$. Given a consistent spatial discretisation on the computational domain $\Omega \subset \mathbb{R}^3$, the Navier--Stokes equations are approximated by a finite-dimensional dynamical system with state vector $\boldsymbol{u}=(\boldsymbol{u}_i)_{i=1}^{N}\in\mathbb{R}^{3N}$, where $\boldsymbol{u}_i\in\mathbb{R}^3$ denotes the velocity at the spatial point $\boldsymbol{x}_i\in\Omega$. The resulting semi-discrete dynamics are written as
\begin{equation}\label{eq:NS}
    \frac{\mathrm{d}\boldsymbol{u}}{\mathrm{d}t}
    = \mathcal{F}(\boldsymbol{u}),
    \qquad
    \boldsymbol{u}_t = \Phi^t(\boldsymbol{u}_0),
\end{equation}
where $\mathcal{F}$ denotes the spatially discretised Navier--Stokes operator, and the flow map $\Phi^t$ maps an initial condition $\boldsymbol{u}_0$ to the corresponding state $\boldsymbol{u}_t$ after time $t$.
The state of this turbulent dynamical system is fully specified by the solenoidal velocity field $\boldsymbol{u}_t$, as it uniquely determines the pressure gradient by the incompressibility constraint and the absolute pressure has no effect on the dynamics \citep{pope_turbulent_2000}.
Although it is possible to represent the velocity field explicitly in a solenoidal basis for a given discretisation, we leave the incompressibility constraint implicit and adopt a discretisation-agnostic approach.

The remainder of this section describes the approximation of the natural distribution of this finite-dimensional dynamical system. 
Although a complete determination of the natural distribution is extremely difficult \citep{monin_statistical_1971}, we elaborate in §\ref{sec:natural_distribution} why probabilistic machine learning models offer a new and promising approach to approximate it from data.
In §\ref{sec:minimal-conditional-flow-units}, we introduce the minimal conditional flow unit on which the machine-learned distribution is trained. 
After presenting the flow-based generative model in §\ref{sec:flow-model} and deriving the conditional sampling algorithm in §\ref{sec:conditional_sampling}, we summarise the sampling procedure with the machine-learned probability distribution and detail the numerical set-up in §\ref{sec:summary_distribution}.

\subsection{Approximating the natural distribution with data}\label{sec:natural_distribution}
The natural distribution of a dynamical system is an invariant probability distribution whose ensemble statistics agree with the long-time statistics of trajectories starting from typical initial conditions \citep{cvitanovic_chaos_2020}. Although the formal construction of such a distribution is extremely difficult for turbulent dynamical systems  \citep{ruelle_hydrodynamic_2012}, we demonstrate that its two essential properties, dynamical invariance and physical representativeness, can be incorporated relatively straightforwardly in a data-driven approach.

The property of dynamical invariance was already considered by \cite{hopf_statistical_1952}, who derived an equation for the evolution of the characteristic functional under the Navier–Stokes dynamics and studied its stationary solutions, corresponding to probability distributions in the phase space of the flow field that remain invariant under those dynamics.
Such an invariant probability measure can be constructed as the time average of Dirac deltas along a trajectory $\boldsymbol{u}_t=\Phi^t(\boldsymbol{u}_0)$ \citep{eckmann_ergodic_1985} 
\begin{equation}\label{eq:invariant_measure}
    \rho(\boldsymbol{u}) = \lim_{T \rightarrow \infty} \frac{1}{T} \int_0^T \delta(\boldsymbol{u}-\boldsymbol{u}_t) ~\mathrm{d} t.
\end{equation}
Hence the infinite-time average of any continuous observable $\varphi(\boldsymbol{u})$ is given by the ensemble average over the invariant probability measure
\begin{equation}\label{eq:invariant_measure_phi}
     \int \varphi(\boldsymbol{u}) ~\mathrm{d} \rho (\boldsymbol{u}) =   \lim_{T \rightarrow \infty} \frac{1}{T}  \int_0^T \varphi (\boldsymbol{u}_t) ~\mathrm{d} t.
\end{equation}
If the invariant probability measure $\rho$ is also indecomposable or ergodic, then the time averages in \eqref{eq:invariant_measure} and \eqref{eq:invariant_measure_phi} reproduce that measure for all initial conditions $\boldsymbol{u}_0$, except possibly for a set of initial conditions whose $\rho$-measure is zero \citep{eckmann_ergodic_1985}. 
However, there are many invariant ergodic measures, so one should choose a measure that is `physical' to describe turbulence \citep{ruelle_what_1978}. For instance, by selecting the laminar solution as initial condition in \eqref{eq:invariant_measure} one could construct an unphysical measure, in the sense that its ensemble statistics differ from those of trajectories that start arbitrarily close to the laminar solution and eventually become turbulent. As the world is intrinsically a little noisy, 
observing exactly the laminar solution has zero probability, whereas observing a perturbation within a certain range of amplitudes is probable, or `typical'. 
To be physically relevant, the measure should thus not only be invariant under the dynamics, but also observable from typical initial conditions.

An invariant measure $\rho(\boldsymbol{u})$ is called physical if there is a positive Lebesgue measure set $V$ in the admissible phase space such that \eqref{eq:invariant_measure_phi} holds for every initial condition $\boldsymbol{u}_0 \in V$ \citep{eckmann_ergodic_1985,young_what_2002}.
This property does not follow from Birkhoff's ergodic theorem, since a physical invariant measure $\rho$ can describe the asymptotic behaviour of trajectories starting outside its support, which additionally has Lebesgue measure zero for a genuinely volume decreasing attractor.
In this sense, $\rho$ is observable not because randomly chosen initial conditions are likely to lie on its support, but because a positive-Lebesgue-measure set of initial conditions, lying in the basins of one or more attractors associated with $\rho$, gives rise to long-time statistics governed by $\rho$.
Another way to define a physical or observable measure was put forth by Kolmogorov, based on the fact that any realistic physical system is also influenced by external noise \citep{eckmann_ergodic_1985,young_what_2002}. The Kolmogorov measure is then defined as the zero-noise limit of the normally unique stationary measure of the noisy dynamical system, which is described by a stochastic process \citep{eckmann_ergodic_1985}.
As any physical turbulent system is subject to external noise, we can coarse-grain the invariant measure from \eqref{eq:invariant_measure}, with an unknown kernel $K_\varepsilon$ with finite noise level $\varepsilon$. In that respect, we may think of the natural distribution in terms of a regularised probability density, obtained by
\begin{subeqnarray}\label{eq:smoothed-distribution}
    \pi (\boldsymbol{u}) &=& \int K_\varepsilon (\boldsymbol{u} - \boldsymbol{v}) ~\mathrm{d} \rho(\boldsymbol{v}) ,\\
    &=& \lim_{T \rightarrow \infty} \frac{1}{T} \int_0^T K_\varepsilon (\boldsymbol{u} - \boldsymbol{u}_t) ~\mathrm{d} t,
\end{subeqnarray}
where the trajectory $\boldsymbol{u}_t=\Phi^t(\boldsymbol{u}_0)$ starts from a `typical' initial condition $\boldsymbol{u}_0$. 
Although physical invariant measures, specifically Sinai--Ruelle--Bowen (SRB) measures, are well understood for uniformly hyperbolic systems and have been constructed for some nonuniformly hyperbolic systems, their rigorous construction for fully developed turbulent flows remains an open problem \citep{young_what_2002, ruelle_hydrodynamic_2012}.

Probabilistic machine-learning models can approximate the natural distribution directly from samples, making it relatively straightforward to incorporate the properties described above. 
The dynamical invariance property can be incorporated implicitly by drawing these samples as snapshots from a sufficiently long trajectory with stationary statistics. Whether these stationary statistics are physical, i.e. representative of typical initial conditions and realisations of the external noise, can then be assessed empirically by comparing them with statistics obtained from independent numerical experiments using a different initial condition or discretisation. 
Moreover, recent probabilistic machine-learning models provide scalable computational methods to approximate the full natural distribution, rather than only selected statistics.

\subsection{Minimal conditional flow unit}\label{sec:minimal-conditional-flow-units}
It is desirable to represent the approximated natural distribution on a minimal domain size and resolution, where the latter can be determined from synchronisation experiments (cf. \citealt{yoshida_regeneration_2005,lalescu_synchronization_2013,wang_synchronization_2022}). 
In DNS, the smallest periodic domain that can sustain the turbulent solution in a channel flow is given by the minimal flow unit \citep{jimenez_minimal_1991,flores_hierarchy_2010}.
However, larger
turbulent structures may emerge in larger domains
\citep{lozano-duran_effect_2014}, and such structures are not necessarily
compatible with exact periodic repetition of the minimal flow unit. 
Therefore, we introduce the minimal \emph{conditional} flow unit, which allows approximate sampling of (non-periodic) flow fields on arbitrarily large domains. 
To construct such a minimal conditional flow unit, we make use of conditional probability, the decay of velocity correlations, and the statistical homogeneity in directions under which the system is translationally invariant.

We define the minimal conditional flow unit as the smallest domain in the homogeneous direction(s) outside which a conditional velocity field, conditioned on a velocity component at the domain centre, is indistinguishable from corresponding unconditional fields in terms of mean-square discrepancy from other conditional realisations.
Let $u_o$ denote an observed or known velocity component at the centre of the unit and $u_u$ an unobserved component outside of it. 
The MSE between a conditional realisation $u_u^c \sim \pi(u_u|u_o)$ and an independent marginal realisation $u_u^m \sim \pi(u_u)$ equals
\begin{equation}\label{eq:mse-marg}
\langle (u_u^{c} - u_u^{m})^2 | u_o \rangle                      = \operatorname{Var}(u_u | u_o) +   \operatorname{Var}(u_u)  + (\langle  u_u | u_o \rangle -  \langle  u_u  \rangle)^2  ,
\end{equation}
which follows from the variance identity. 
For two independent conditional realisations $u_u^{c}$ and $u_u^{c,\mathrm{new}} \sim \pi(u_u|u_o)$, the MSE is given by
\begin{equation}\label{eq:mse-cond}
  \langle (u_u^{c} - u_u^{c,\mathrm{new}})^2 | u_o \rangle   =  2 \operatorname{Var}(u_u | u_o).
\end{equation}
To quantify the difference in MSE, we approximate the conditional mean  $ \langle u_u | u_o\rangle$ and conditional variance $\operatorname{Var}(u_u | u_o) $ as in linear stochastic estimation \citep{adrian_stochastic_1996} as
\begin{subeqnarray}\label{eq:cond_var}
   \langle u_u | u_o\rangle &\approx&  \langle u_u \rangle + \rho_{uo}  (u_o - \langle u_o \rangle )  \sqrt{{\operatorname{Var}(u_u)}/{\operatorname{Var}(u_o)}} , \\
    \operatorname{Var}(u_u | u_o)     &\approx&     \operatorname{Var}(u_u) (1-\rho_{uo}^2), 
\end{subeqnarray}
where $\rho_{uo}$ is the cross-correlation between $u_u$ and $u_o$. 
Hence, we find that
\begin{equation}
\langle (u_u^{c} - u_u^{m})^2 | u_o \rangle -\langle (u_u^{c} - u_u^{c,\mathrm{new}})^2 | u_o \rangle   \approx  \rho_{uo}^2  \operatorname{Var}(u_u) \left( \frac{ (u_o - \langle u_o \rangle )^2}{\operatorname{Var}(u_o)} + 1 \right) .
\end{equation}
When the squared cross-correlation $\rho_{uo}^2$ is sufficiently small, the difference between the MSEs involving marginal and conditional samples is negligible, as both MSEs then approach $2\operatorname{Var}(u_u)$. Similarly, it can be shown that the MSEs of conditional and marginal samples with respect to the conditional mean both approach $\operatorname{Var}(u_u)$. 
In this limit, the conditional and unconditional fields are practically indistinguishable in terms of their MSE relative to other conditional realisations. The same conclusion holds relative to the conditionally averaged velocity field, which represents the conditional eddy associated with $u_o$ in linear stochastic estimation \citep{adrian_stochastic_1996}.

\begin{figure}
    \centering   
    \includegraphics[width=\linewidth]{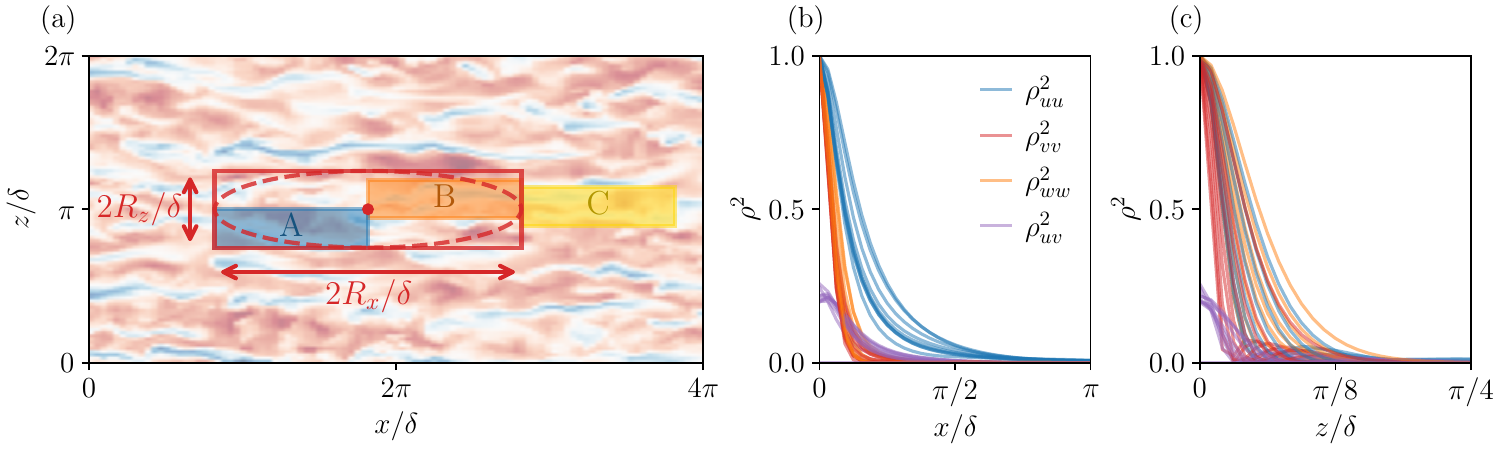}

    \caption{Definition of the minimal conditional flow unit in a turbulent channel flow: (a) the minimal conditional flow unit in red, (b) squared streamwise correlation at different wall-normal heights, (c) squared spanwise correlation at different wall-normal heights. The correlations are taken from the database by \cite{kim_turbulence_1987}. In (a) the subdomains $A$, $B$, and $C$ used in the sequential conditional sampling argument are also shown.}
    \label{fig:correlations}
\end{figure}

The dimensions of the minimal conditional flow unit in a given turbulent flow can hence be estimated from its two-point correlation structure.
Let $R_h$ denote the distance in a homogeneous direction $h$ beyond which all relevant squared cross-correlations are sufficiently small. Then the minimal conditional flow unit spans $2R_h$ in that homogeneous direction, since a velocity component at the centre of the unit must be separated by at least $R_h$ from components outside the unit on either side. 
Figure~\ref{fig:correlations}(a) shows the minimal conditional flow unit in a turbulent channel flow. As only the wall-normal direction is inhomogeneous, the minimal conditional channel-flow unit has size $2R_x \times 2\delta \times 2R_z$, where $\delta$ is the channel half-height, $x$ represents the streamwise direction, and $z$ the spanwise direction. 
From the squared cross-correlations computed from the database of \cite{kim_turbulence_1987} and shown in figure~\ref{fig:correlations}(b,c), we estimate $R_x \approx \upi\delta$ and $R_z \approx \upi\delta/4$, giving approximate dimensions $2\upi\delta \times 2\delta \times \upi\delta/2$ to the minimal conditional channel flow unit.

When the natural distribution is represented on the minimal conditional flow unit, flow fields on larger domains can still be sampled approximately through sequential conditional sampling.
Consider three subdomains $A$, $B$, and $C$ that are adjacent in a homogeneous direction. Each subdomain has width $R_x$ and $R_z$ in the homogeneous directions $x$ and $z$, and spans the full domain in the inhomogeneous direction, as shown in figure~\ref{fig:correlations}(a).
Let $\boldsymbol{{u}}_A = (\boldsymbol{u}_i)_{\boldsymbol{x}_i\in A}$ denote the discretised velocity field restricted to subdomain $A$, with
$\boldsymbol{u}_B$ and $\boldsymbol{u}_C$ defined analogously.
The natural distribution of the flow field on the combined domain $A \cup  B \cup C$ can then be decomposed using the chain rule of probability: 
\begin{equation}\label{eq:cond_decomposition}
    \pi (\boldsymbol{{u}}_A, \boldsymbol{{u}}_B,\boldsymbol{{u}}_C) = \pi (\boldsymbol{{u}}_A) \pi ( \boldsymbol{{u}}_B | \boldsymbol{{u}}_A) \pi (\boldsymbol{{u}}_C | \boldsymbol{{u}}_A, \boldsymbol{{u}}_B).
\end{equation}
This decomposition suggests a sequential sampling procedure: first sample $\boldsymbol{u}_A$, then sample $\boldsymbol{u}_B$ conditionally on $\boldsymbol{u}_A$, and finally sample $\boldsymbol{u}_C$ conditionally on $\boldsymbol{u}_A$ and $\boldsymbol{u}_B$.  
However, the minimal conditional flow unit contains information only over an extent of $2R_h$ in every homogeneous direction. It therefore provides access to local conditional distributions such as $\pi(\boldsymbol{u}_B|\boldsymbol{u}_A)$ and
$\pi(\boldsymbol{u}_C|\boldsymbol{u}_B)$, but not directly to the full three-subdomain conditional $\pi(\boldsymbol{u}_C|\boldsymbol{u}_A,\boldsymbol{u}_B)$, which spans an extent of $3R_h$.
We therefore introduce the local-conditioning approximation
\begin{equation}\label{eq:local_conditioning}
    \pi(\boldsymbol{u}_C | \boldsymbol{u}_A,\boldsymbol{u}_B)
    \approx
    \pi(\boldsymbol{u}_C | \boldsymbol{u}_B).
\end{equation}
By statistical homogeneity, the local conditional distributions are invariant under translations in the homogeneous direction, so that $\pi_{C| B}=\pi_{B| A}$. Under this approximation, flow fields on larger domains $A\cup\cdots\cup Z$ can be generated by sequentially sampling new subdomains, each conditioned only on the previously sampled neighbouring subdomain:
\begin{equation}\label{eq:cond_decomposition_approx}
    \pi(\boldsymbol{u}_A,\ldots,\boldsymbol{u}_Z)
    \approx
    \pi(\boldsymbol{u}_A)
    \pi(\boldsymbol{u}_B | \boldsymbol{u}_A)
    \cdots
    \pi(\boldsymbol{u}_Z | \boldsymbol{u}_Y).
\end{equation}
Note that the sequential local conditioning procedure does not rule out long‑range dependencies, as these can be propagated indirectly through previously generated subdomains.

The validity of the local-conditioning approximation is not guaranteed by marginal decorrelation alone. In particular, even if $\boldsymbol{u}_A$ and $\boldsymbol{u}_C$ are uncorrelated, this does not imply that they are conditionally independent given $\boldsymbol{u}_B$.\footnote{For example, let $u_A \sim \mathcal{N}(0,1)$, $u_C \sim \mathcal{N}(0,1)$ be independent, and define $u_B=(u_A+u_C)/\sqrt{2}$. Then $\rho_{AC}=0$, while $\rho_{AB}=\rho_{BC}=1/\sqrt{2}$ and $\sigma_A^2=\sigma_B^2=\sigma_C^2=1$. Hence $u_C| u_A \sim u_C \sim \mathcal{N}(0,1)$ and $u_C| u_B \sim \mathcal{N}(u_B/\sqrt{2},1/2)$, whereas $\pi(u_C| u_A,u_B)=\delta(u_A+u_C-\sqrt{2}u_B)$ is deterministic, which is not obvious from the correlation structure alone.}
Information from $A$ may still be transmitted to $C$ through $B$ when both regions are correlated with $B$, a mechanism known as the relay effect \citep{chiles_geostatistics_2012}. Consequently, even choosing $R_h$ as the distance at which correlations vanish exactly does not guarantee that conditioning only on $B$ is equivalent to conditioning on both $A$ and $B$. 
Nevertheless, distant information often contributes little beyond nearby information because of the screening effect, which is known to be strong for certain covariance structures \citep{stein_screening_2002,stein_2010_2011}. 
Motivated by this screening approximation, we assume that the field in region $B$ contains most, or at least sufficient, information about the field in region $C$. 
This assumption is natural in the present construction because $B$ contains the neighbouring points that exert a non-negligible influence on $C$ according to the MSE-criterion, whereas points in $A$ are farther away than $R_h$ in the homogeneous direction. Since the minimal conditional flow unit spans $2R_h$, conditioning on a region of width $R_h$ allows one to sample the velocity field up to an additional distance $R_h$ in the homogeneous direction. 
This motivates the subdomain decomposition in \eqref{eq:cond_decomposition_approx} and provides the basis for sequential conditional sampling of flow fields on larger domains, discussed in §\ref{sec:summary_distribution}.

\subsection{Flow-based generative model}\label{sec:flow-model}
Flow-based generative models learn a transformation of a prescribed probability distribution, typically a standard Gaussian, to a target probability distribution.
In our case, the target is the coarse-grained, or smoothed, natural distribution of the velocity field in \eqref{eq:smoothed-distribution}, restricted to a minimal conditional flow unit. From this point onward, we denote by $\boldsymbol{u}\in\mathbb{R}^n$ the discretised velocity field restricted to such a unit, where $n$ is the number of remaining degrees of freedom, and write its probability density as $\pi (\boldsymbol{u})$.

In continuous normalising flows \citep{chen_neural_2019}, a dynamical system for the evolution of a state $\boldsymbol{v}\in \mathbb{R}^{ n}$ in terms of a fictitious time coordinate $\xi \in [0,1)$ is introduced, i.e.,
\begin{equation}\label{eq:ODE}
    \frac{\mathrm{d} \boldsymbol{v}}{\mathrm{d} \xi} = \boldsymbol{f} (\boldsymbol{v},\xi), \qquad \boldsymbol{v}_{\! \xi}= \Psi^\xi(\boldsymbol{v}_0),
\end{equation}
through a $\xi$-dependent vector field $\boldsymbol{f}(\boldsymbol{v},\xi)$.
Although this vector field is commonly termed the velocity field in the flow-matching literature \citep{lipman_flow_2024,lai_principles_2025}, we refer to it as the generator to distinguish it from the physical velocity field.
The flow map $\Psi^\xi$ defined by this set of ordinary differential equations deterministically maps an initial condition $\boldsymbol{v}_0$ to its state $\boldsymbol{v}_{\! \xi}$ at time $\xi$. 
If the initial condition $\boldsymbol{v}_0$ is drawn from a prescribed base distribution $p_0$, then the flow map induces a time-dependent probability density or probability path $(p_{\! \xi})_{0\leq \xi < 1}$, with $\boldsymbol{v}_{\! \xi}\sim p_{\! \xi}(\cdot)$.
For sufficiently regular generators, the flow map is invertible and probability mass is conserved along trajectories \citep{villani_optimal_2009,chen_neural_2019}. The density $p_{\! \xi}$ therefore satisfies the probability continuity equation
\begin{equation}\label{eq:continuity}
   \frac{\partial }{\partial \xi } p_{\! \xi}(\boldsymbol{v})+ \nabla_{\boldsymbol{v}}  \cdot [\boldsymbol{f} (\boldsymbol{v}, \xi) p_{\! \xi}(\boldsymbol{v})]=0.
\end{equation}
The goal is then to find a generator $\boldsymbol{f} (\boldsymbol{v}, \xi)$ whose induced terminal density  $p_{\! \xi \rightarrow1}\equiv p_{1}$ satisfies $p_1\approx \pi$, where $\pi$ is the target distribution. 
Once such a generator has been found, synthetic turbulent velocity fields can be generated by drawing $\boldsymbol{v}_0\sim p_0(\cdot)$ and integrating \eqref{eq:ODE} from $\xi=0$ to $\xi=1$, understood in the improper sense. 
Note that the dynamical system from \eqref{eq:NS} governed by the Navier--Stokes dynamics essentially accomplishes the same task during spin-up of turbulence, so the true objective is finding an alternative generator and probability path that achieve this goal at a much lower computational cost.

The question thus becomes how to effectively couple the initial distribution $p_0$ to the target distribution $\pi$ through a probability path $p_{\! \xi}$, induced by a generator $\boldsymbol{f}$ in the sense of \eqref{eq:ODE} and \eqref{eq:continuity}. 
In the flow-matching paradigm \citep{lipman_flow_2024}, such a coupling is constructed by specifying a conditional probability path $p_{\! \xi}(\boldsymbol{v}|\boldsymbol{u})$ for all samples from the target distribution $\boldsymbol{u}  \sim \pi(\cdot)$. 
After marginalisation, the corresponding (marginal) probability path is then obtained for $\xi \in [0,1)$ as 
\begin{equation}\label{eq:marginalization_trick}
   p_{\! \xi}(\boldsymbol{v})=\int p_{\! \xi}(\boldsymbol{v} |\boldsymbol{u})\pi(\boldsymbol{u}) \mathrm{d} \boldsymbol{u}.
\end{equation}
As a result, the requirement that the final distribution $p_1$ equals the target distribution $ \pi $ is satisfied when the conditional probability path converges to the Dirac delta distribution $p_1(\boldsymbol{v}|\boldsymbol{u}) = \delta(\boldsymbol{v} - \boldsymbol{u})$. For the initial distribution $p_0 $, a standard Gaussian can be chosen with $p_0(\boldsymbol{v} |\boldsymbol{u}) = \mathcal{N} (\boldsymbol{v}; \boldsymbol{0}, \mathsfbi{I})$. 

Let the conditional path $p_{\! \xi}(\boldsymbol{v}|\boldsymbol{u})$ satisfy the probability continuity equation with conditional generator $\boldsymbol{f}(\boldsymbol{v},\xi|\boldsymbol{u})$ for each target sample $\boldsymbol{u}$ and for $\xi \in [0,1)$, where the open interval allows for target distributions with compact support \citep{lipman_flow_2024}.
Then the marginal path $p_{\! \xi}(\boldsymbol{v})$, defined in \eqref{eq:marginalization_trick}, satisfies the probability continuity equation for $\xi \in [0,1)$ with the marginal generator \citep{lipman_flow_2023}
\begin{equation}\label{eq:marginal_generator}
    \boldsymbol{f}(\boldsymbol{v},\xi)    =    \int    \boldsymbol{f}(\boldsymbol{v},\xi|\boldsymbol{u}) 
    \frac{
        p_{\! \xi}(\boldsymbol{v}|\boldsymbol{u})\pi(\boldsymbol{u})
    }{
        p_{\! \xi}(\boldsymbol{v})
    }
    \,\mathrm{d}\boldsymbol{u}.
\end{equation}
In practice, this integral need not be evaluated explicitly. Instead, a neural network $\boldsymbol{f}_{\!\boldsymbol{\vartheta}}$ with parameters $\boldsymbol{\vartheta}$ is trained to approximate $\boldsymbol{f}$, based on a chosen conditional probability path.
One approach is to select the conditional path as the minimiser of a cost functional with desirable properties. The path minimising the 2‑norm of the conditional generator along its associated conditional probability path is known as the Gaussian conditional optimal transport path \citep{lipman_flow_2024}.
It is given by the conditional flow map $ \Psi^\xi(\boldsymbol{v}_0  \vert \boldsymbol{u}) = \xi \boldsymbol{u} +(1-\xi) \boldsymbol{v}_0, $ with as conditional generator and probability path 
\begin{subeqnarray}\label{eq:condot-generator}
    \boldsymbol{f}(\boldsymbol{v}, \xi  \vert \boldsymbol{u}) &=& \frac{\boldsymbol{u}-\boldsymbol{v}}{1-\xi} ,\\
    p_{\! \xi}(\boldsymbol{v} | \boldsymbol{u}) &=& \mathcal{N}(\boldsymbol{v}; \xi \boldsymbol{u}, (1-\xi)^2 \mathsfbi{I}).
\end{subeqnarray}
By construction, this conditional path yields straight trajectories in phase space, which has been empirically associated with faster training and sampling, as well as better generalisation, compared to standard diffusion paths \citep{lipman_flow_2023}. 
Moreover, the forward Euler and standard fourth-order Runge--Kutta methods are stable for this conditional generator on any equidistant grid in $[0,1)$ and reproduce its affine trajectories exactly in exact arithmetic.

The target probability path is matched by training the neural network $\boldsymbol{f}_{\!\boldsymbol{\vartheta}}(\boldsymbol{v}, \xi)$ to approximate the target generator $\boldsymbol{f}(\boldsymbol{v}, \xi)$ along $p_{\! \xi}(\boldsymbol{v})$ through the flow-matching loss \citep{lipman_flow_2023}:
\begin{equation}\label{eq:FM}
    \mathcal{L}_{\textit{FM}}(\boldsymbol{\vartheta}) 
    = \mathbb{E}_{\xi \sim \mathcal{U}_0^1,\, \boldsymbol{v} \sim p_{\! \xi}(\cdot)}
    \left[
    \left\| 
    \boldsymbol{f}_{\!\boldsymbol{\vartheta}}(\boldsymbol{v}, \xi) 
    - \boldsymbol{f}(\boldsymbol{v}, \xi)
    \right\|_2^2
    \right].
\end{equation}
Unfortunately, the flow-matching loss is intractable due to the integration over the entire target distribution in the definition of the target generator in \eqref{eq:marginal_generator}. However, \cite{lipman_flow_2023} showed that the gradient of the flow-matching loss with respect to the neural network parameters is equal to that of the conditional flow-matching loss
\begin{subeqnarray}\label{eq:CFM}
\mathcal{L}_{\textit{CFM}}(\boldsymbol{\vartheta}) =  &&\mathbb{E}_{\xi \sim \mathcal{U}_0^1,\boldsymbol{u}\sim \pi(\cdot), \boldsymbol{v} \sim p_{\! \xi}(\cdot \vert \boldsymbol{u})} \left[ \| \boldsymbol{f}_{\!\boldsymbol{\vartheta}} \left( \boldsymbol{v},\xi \right) - \boldsymbol{f}(\boldsymbol{v}, \xi \vert \boldsymbol{u})\|_2^2 \right],  \\
=  &&\mathbb{E}_{\xi \sim \mathcal{U}_0^1, \boldsymbol{u}\sim \pi(\cdot), \boldsymbol{v}_0 \sim p_0(\cdot) } \left[ \| \boldsymbol{f}_{\!\boldsymbol{\vartheta}} \left(\Psi^\xi(\boldsymbol{v}_0  \vert \boldsymbol{u}),\xi \right) - (\boldsymbol{u}-\boldsymbol{v}_0)\|_2^2 \right],
\end{subeqnarray}
where the second equality is obtained for the chosen Gaussian conditional optimal transport path. 
Consequently, the neural-network generator can be trained to match the generator of straight-line trajectories connecting initial states drawn from the base distribution $p_0$ to states drawn from the target distribution $\pi$.
The learned generator $\boldsymbol{f}_{\!\boldsymbol{\vartheta}}^\star \left( \boldsymbol{v},\xi \right)$ can then be used to generate synthetic turbulent velocity fields that approximately follow the natural distribution, by integrating the set of ordinary differential equations of \eqref{eq:ODE} with $\boldsymbol{f} \left( \boldsymbol{v},\xi \right) \approx \boldsymbol{f}_{\!\boldsymbol{\vartheta}}^\star \left( \boldsymbol{v},\xi \right)$ and initial conditions drawn from a standard Gaussian.

\subsection{Approximate sampling of the conditional natural distribution}\label{sec:conditional_sampling}
We now introduce a modification to the flow-based generative model, so that we can also approximately sample the conditional natural distribution.
To this end, we first derive the training loss for a guided flow-based model that transports a standard Gaussian to the conditional natural distribution. We then construct an approximate minimiser of this loss by relating the guided generator to the previously learned marginal generator (see §\ref{sec:flow-model}).

Consider the conditional natural distribution $\pi(\boldsymbol{u}_u \vert \boldsymbol{u}_o)$. 
Here, $\boldsymbol{u}_o=\mathsfbi{M}\boldsymbol{u}$ denotes an observed, or known, part of the velocity field. The mask matrix $\mathsfbi{M} \in \mathbb{R}^{m \times n}$ contains $m$ distinct rows of the identity matrix and therefore selects the observed components.
The unobserved part of the velocity field is denoted by $\boldsymbol{u}_u=\mathsfbi{M}^\perp \boldsymbol{u}$. The complementary mask $\mathsfbi{M}^\perp \in \mathbb{R}^{(n-m) \times n}$ contains the remaining $n-m$ rows of the identity matrix.
With the conditional natural distribution as target distribution, we can construct a guided generator $ \boldsymbol{f}(\boldsymbol{v}, \xi \vert \boldsymbol{u}_o)$ and corresponding guided probability path $ p_{\! \xi}(\boldsymbol{v} \vert \boldsymbol{u}_o )$ for $\boldsymbol{v}\in \mathbb{R}^n$,
\begin{subeqnarray}
   p_{\! \xi}(\boldsymbol{v} \vert \boldsymbol{u}_o )&=&\int p_{\! \xi}(\boldsymbol{v} |\boldsymbol{u})\pi(\boldsymbol{u}_u \vert \boldsymbol{u}_o) \mathrm{d} \boldsymbol{u}_u, \\
        \boldsymbol{f}(\boldsymbol{v}, \xi \vert \boldsymbol{u}_o) &=& \int  \boldsymbol{f}(\boldsymbol{v}, \xi | \boldsymbol{u})\frac{p_{\! \xi}(\boldsymbol{v} |\boldsymbol{u})\pi(\boldsymbol{u}_u \vert \boldsymbol{u}_o)}{p_{\! \xi}(\boldsymbol{v} \vert \boldsymbol{u}_o )} \mathrm{d}\boldsymbol{u}_u,
\end{subeqnarray}
that satisfy the probability continuity equation for $\xi \in [0,1)$, if the conditional generator $\boldsymbol{f}(\boldsymbol{v}, \xi | \boldsymbol{u})\equiv \boldsymbol{f}(\boldsymbol{v}, \xi | \boldsymbol{u}_u, \boldsymbol{u}_o) $ and conditional probability path $p_{\! \xi}(\boldsymbol{v} |\boldsymbol{u})\equiv p_{\! \xi}( \boldsymbol{v} | \boldsymbol{u}_u, \boldsymbol{u}_o) $ do so.

The neural network for the guided generator $\boldsymbol{f}_{\boldsymbol{ \vartheta}}(\boldsymbol{v},\xi | \boldsymbol{u}_o)$ should then be trained by minimising its discrepancy with the target generator $\boldsymbol{f}(\boldsymbol{v},\xi | \boldsymbol{u}_o)$ over the target probability path $ p_{\! \xi}(\boldsymbol{v} \vert \boldsymbol{u}_o )$ and for all realisations of the observed part of the state $\boldsymbol{u}_o \sim \pi_o(\cdot)$, through the guided flow-matching loss 
\begin{equation} 
    \mathcal{L}_{GFM} (\boldsymbol{\vartheta}) = \mathbb{E}_{\xi \sim \mathcal{U}_0^1, \boldsymbol{u}_o \sim \pi_o(\cdot), \boldsymbol{v} \sim p_\xi(\cdot | \boldsymbol{u}_o)} [\| \boldsymbol{f}_{\boldsymbol{\vartheta}} (\boldsymbol{v}, \xi | \boldsymbol{u}_o) - \boldsymbol{f}(\boldsymbol{v}, \xi | \boldsymbol{u}_o) \|^2] .
\end{equation}
Similar to before, it can be shown that the gradient of this loss with respect to the neural network parameters $\boldsymbol{\vartheta}$ equals that of a guided conditional flow-matching loss
\begin{subeqnarray}\label{eq:GCFM}
    \mathcal{L}_\textit{GCFM}(\boldsymbol{\vartheta})  =\mathbb{E}_{\xi \sim \mathcal{U}_0^1, \boldsymbol{u} \sim \pi(\cdot), \boldsymbol{v} \sim p_\xi(\cdot | \boldsymbol{u})} [\| \boldsymbol{f}_{\boldsymbol{\vartheta}} (\boldsymbol{v}, \xi | \boldsymbol{u}_o) - \boldsymbol{f}(\boldsymbol{v}, \xi | \boldsymbol{u}) \|^2], \\
   =   \mathbb{E}_{\xi \sim \mathcal{U}_0^1, \boldsymbol{u} \sim \pi(\cdot ), \boldsymbol{v}_0 \sim p_0(\cdot) } [\| \boldsymbol{f}_{\!\boldsymbol{\vartheta}} \left(\Psi^\xi(\boldsymbol{v}_0  \vert \boldsymbol{u}),\xi  \vert \boldsymbol{u}_o \right) - (\boldsymbol{u}-\boldsymbol{v}_0) \|^2] ,
\end{subeqnarray}
where $\pi(\boldsymbol{u})=\pi_o(\boldsymbol{u}_o)\pi(\boldsymbol{u}_u|\boldsymbol{u}_o)$ and the second equality is obtained for the conditional optimal transport path.  
Consequently, the conditional natural distribution can be approximately sampled by integrating the guided generator $\boldsymbol{f}^\star_{\boldsymbol{ \vartheta}}(\boldsymbol{v},\xi | \boldsymbol{u}_o)$ that minimises the guided conditional flow-matching loss, starting from states drawn from the initial distribution $\boldsymbol{v}_0 \sim \mathcal{N}(\boldsymbol{0}, \mathsfbi{I})$.

The conditional natural distribution $\pi(\boldsymbol{u}_u \vert \boldsymbol{u}_o)$ can be defined for any mask $\mathsfbi{M}$. However, training a separate guided generator for each possible mask would be impractical. We therefore construct an approximate minimiser of the guided conditional flow-matching loss as follows.
Since the guided generator can take the observed part of the state $\boldsymbol{u}_o$ as input, we can exactly match the target generator for the observed part by choosing
\begin{equation}
    \mathsfbi{M}  \boldsymbol{f}_{\!\boldsymbol{\vartheta}}^\star \left( \boldsymbol{v},\xi \vert \boldsymbol{u}_o \right) = \frac{\boldsymbol{u}_o - \mathsfbi{M} \boldsymbol{v}}{1-\xi}, 
\end{equation}
for the conditional optimal transport path, see (\ref{eq:condot-generator}a). 
This leads to the following solution for the observed part of the state $\mathsfbi{M}\boldsymbol{v}_{\! \xi}= \xi \boldsymbol{u}_o + (1-\xi) \mathsfbi{M}\boldsymbol{v}_0$.
As a result, only the components of the generator selected by $\mathsfbi{M}^\perp$ contribute to the guided conditional flow-matching loss in (\ref{eq:GCFM}b). 
As the guided generator should be usable with any mask $\mathsfbi{M}$, we average the loss over all possible masks $\mathsfbi{M}\in \mathbb{R}^{m \times n}$ containing $m$ random, distinct rows of the identity matrix with $m \in \{1, \ldots, n-1 \}$, i.e.,
\begin{equation}\label{eq:masked-GCFM}
    \mathbb{E}_{\mathsfbi{M},\xi \sim \mathcal{U}_0^1, \boldsymbol{u} \sim \pi(\cdot ), \boldsymbol{v}_0 \sim p_0(\cdot) } [\| \mathsfbi{M}^\perp  \left( \boldsymbol{f}_{\!\boldsymbol{\vartheta}} \left(\Psi^\xi(\boldsymbol{v}_0  \vert \boldsymbol{u}),\xi  \vert \boldsymbol{u}_o \right) - (\boldsymbol{u}-\boldsymbol{v}_0) \right) \|^2] ,
\end{equation}
Here, $\mathsfbi{M}^\perp$ explicitly selects the unobserved components, for which the discrepancy is not necessarily zero.
If the unobserved part of the guided generator network $\mathsfbi{M}^\perp \boldsymbol{f}_{\!\boldsymbol{\vartheta}}(\boldsymbol{v},\xi \vert \boldsymbol{u}_o )$ is restricted to only depend on $\boldsymbol{u}_o$ through $\mathsfbi{M}\boldsymbol{v}_{\! \xi}  = \xi \boldsymbol{u}_o + (1-\xi) \mathsfbi{M}\boldsymbol{v}_0$ for all $\xi \in [0,1)$, then \eqref{eq:masked-GCFM} becomes a random selection of $n-m$ component-wise squared errors from the conditional flow-matching loss in (\ref{eq:CFM}a). Under this restriction and under a mask distribution that weights all components equally, the previously trained generator can therefore be reused as an approximate minimiser of \eqref{eq:masked-GCFM}, so that $\mathsfbi{M}^\perp \boldsymbol{f}_{\!\boldsymbol{\vartheta}}^\star(\boldsymbol{v},\xi \vert \boldsymbol{u}_o ) \approx \mathsfbi{M}^\perp \boldsymbol{f}_{\!\boldsymbol{\vartheta}}^\star(\boldsymbol{v},\xi) $ if  $\mathsfbi{M}\boldsymbol{v}_{\! \xi}  = \xi \boldsymbol{u}_o + (1-\xi) \mathsfbi{M}\boldsymbol{v}_0$.

Consequently, we can approximately sample the conditional natural distribution $\pi(\boldsymbol{u}_u \vert \boldsymbol{u}_o)$ by integrating the modified system of ordinary differential equations
\begin{equation}\label{eq:conditional_ode}
        \frac{\mathrm{d} \boldsymbol{v}}{\mathrm{d} \xi} 
        =  \mathsfbi{M}^\top \left( \frac{\boldsymbol{u}_o - \mathsfbi{M} \boldsymbol{v}}{1-\xi} \right) + \left( \mathsfbi{M}^\perp \right)^\top  \boldsymbol{\mathsfbi{M}}^\perp \boldsymbol{f}_{\!\boldsymbol{\vartheta}}^\star (\boldsymbol{v},\xi)
\end{equation}
over $\xi \in [0,1)$, starting from an initial condition $\boldsymbol{v}_0\sim \mathcal{N}(\boldsymbol{0}, \mathsfbi{I})$.
Projecting \eqref{eq:conditional_ode} onto the observed and unobserved parts with $\mathsfbi{M}$ and  $\mathsfbi{M}^\perp$ gives the system that is solved in practice:
\begin{subeqnarray}
    \mathsfbi{M} \boldsymbol{v}_{\! \xi} &=& \xi \boldsymbol{u}_o + (1-\xi) \mathsfbi{M} \boldsymbol{v}_0, \\
     \mathsfbi{M}^\perp \frac{\mathrm{d} \boldsymbol{v}_{\! \xi} }{\mathrm{d} \xi} &=&  \mathsfbi{M}^\perp \boldsymbol{f}_{\!\boldsymbol{\vartheta}}^\star (\boldsymbol{v}_{\! \xi},\xi).
\end{subeqnarray}
Note that, owing to the approximation introduced to enable reuse of the unconditional generator, the proposed method does not sample the conditional distribution exactly. This limitation is shared by many practical approaches to conditional generation and inverse problems that rely on a pretrained generative model, including RePaint \citep{lugmayr_repaint_2022} and diffusion posterior sampling \citep{chung_diffusion_2023}, and it remains unclear which of these methods performs best across problem settings \citep{daras_survey_2024}. Many such approaches also introduce one or more tuning parameters, whose selection can substantially affect sampling accuracy and may require problem-dependent calibration. The present approach introduces no additional tuning parameters. Moreover, the resulting system of ODEs can be integrated using the same numerical scheme as for the unconditional system.

\subsection{Summary and case set-up}\label{sec:summary_distribution}

\begin{figure}
    \centering
    \includegraphics[width=\textwidth]{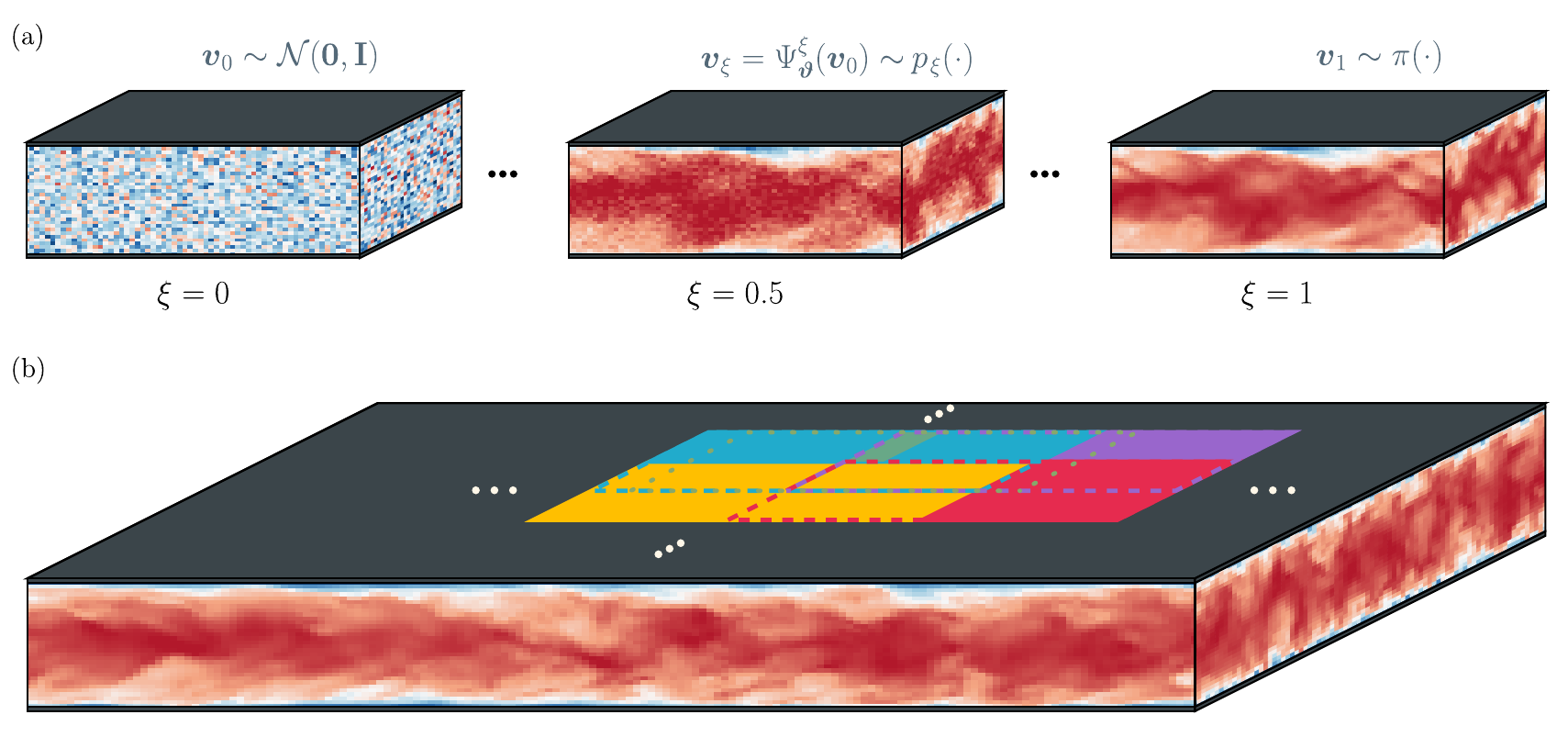}
	\caption{Schematic representation of the sampling process with the flow-based generative model: (a) The flow map $\boldsymbol{v}_{\! \xi}=\Psi^\xi_{\boldsymbol{\vartheta}}(\boldsymbol{v}_0)$ deterministically maps an initial condition $\boldsymbol{v}_0$, sampled from a standard Gaussian, to a turbulent channel flow field $\boldsymbol{v}_{1}$ at fictitious time $\xi=1$. (b) Through sequential conditional sampling with minimal conditional flow units, turbulent channel flow fields on arbitrarily large domains can be generated: (orange) seed unit, (red) streamwise extension, (blue) spanwise extension, (green) incremental extension, (purple) fast corner extension.} 
    \label{fig:generation}
\end{figure}

The sampling procedure of the machine-learned probability distribution in the phase space of turbulent channel flow fields is depicted in figure~\ref{fig:generation}. Since the flow-based generative model is trained on a minimal conditional flow unit, synthetic turbulent velocity fields on such a unit can be generated by integrating the learned set of ordinary differential equations, 
\begin{equation}\label{eq:trained_generator}
    \frac{\mathrm{d} \boldsymbol{v}}{\mathrm{d} \xi}   = \boldsymbol{f}_{\!\boldsymbol{\vartheta}}^\star(\boldsymbol{v},\xi),
\end{equation}
from an initial condition $\boldsymbol{v}_0 \sim \mathcal{N}(\boldsymbol{0}, \mathsfbi{I})$. As shown in figure~\ref{fig:generation}(a), the deterministic flow map $\boldsymbol{v}_{\! \xi}=\Psi_{\boldsymbol{\vartheta}}^\xi(\boldsymbol{v}_0)$ transforms the initial standard Gaussian field into a synthetic turbulent velocity field at $\xi=1$. 
Samples from the conditional natural distribution are obtained in the same way, but by integrating the modified dynamical system from \eqref{eq:conditional_ode}, again starting from a standard Gaussian initial condition. This conditional sampling procedure can be used to reconstruct turbulent velocity fields when part of the velocity field is known. It also enables the generation of synthetic turbulent velocity fields on arbitrarily large domains.

Figure~\ref{fig:generation}(b) schematically depicts this domain-extension procedure. 
First, one minimal conditional flow unit is sampled unconditionally (orange shaded). 
The domain is then extended in the $x$-direction by conditionally generating half a neighbouring unit (red shaded) using the information from the orange region that directly influences it (red dashed), i.e. all available information within a minimal conditional flow unit centred at any $(x,z)$-position in the region to be generated  (cf. §\ref{sec:minimal-conditional-flow-units}).
The same procedure is applied in the $z$-direction (blue). 
In principle, only the upper-left quadrant of the blue unit can be retained directly, since only this region was generated using all available information within surrounding minimal conditional flow units. 
Indeed, the upper-right quadrant of the blue unit depends on the upper quadrant of the red unit and therefore requires a sequential generation procedure. This can be achieved by repeatedly shifting the conditional flow unit by $\Delta$ in the $x$-direction and generating slices of width $\Delta$, each conditioned on all available information within a minimal conditional flow unit (green). Choosing $\Delta$ equal to the grid spacing ensures that every generated slice satisfies this condition. In practice, however, we set $\Delta=R_x$ for a minimal conditional flow unit of $2R_x \times 2\delta \times 2R_z$. This allows the entire blue region to be retained and the purple region to be generated subsequently without discarding intermediate slices, resulting in a substantially more efficient procedure. Independent of whether the domain is tiled in $x$-first or $z$-first order, this choice implies approximating $p(\text{blue, red}|\text{orange})\approx p(\text{blue}|\text{orange}) p(\text{red}|\text{orange})$.
Repeating this process, while shifting the minimal conditional flow units and mirroring the conditioning regions, allows assembling arbitrarily large synthetic turbulent fields. The computational cost of this procedure scales linearly with domain size, while the extensions into the four quadrants can be performed in parallel. In the tests presented in §\ref{sec:large-dom}, no influence of either the tiling order or the seed-unit position on the large-domain statistics is observed.

The reference direct numerical simulation, whose snapshots are used to train the machine-learned probability distribution, is performed with the pseudospectral code \textit{SP-Wind} 
\citep{meyers_is_2007, meyers_les_2010, bon_dns_2022}. 
The flow is governed by the incompressible Navier–Stokes equations on a domain of size $8\upi \delta \times 2 \delta \times 3 \upi \delta$ along the streamwise ($x$), wall-normal ($y$), and spanwise ($z$) directions, where $\delta = 1$ is the half-channel height. The domain is periodic in the streamwise and spanwise directions, with a grid resolution of $\Delta x^+ \times \Delta z^+  = 11.8  \times 5.9$ or $N_x \times N_z =384  \times 288$. 
In the wall-normal direction, a fourth-order energy conservative scheme \citep{verstappen_symmetry-preserving_2003} is used on a grid resolution of $\Delta y^+ \in [0.17, 3.1]$ obtained for $N_y = 256$ with a hyperbolic tangent stretching function. 
The time-integration uses a fourth-order Runge--Kutta method, for which the timestep is chosen such that the Courant number is maintained below 0.4. The flow field is initialised with a random perturbation of the laminar profile and simulated for $40\delta/u_\tau$ to reach statistical stationarity, where $u_\tau$ is the friction velocity. Subsequently, velocity field snapshots are recorded every $\Delta t u_\tau/\delta=0.25$ over an additional $45\delta/u_\tau$. The ensemble statistics of this reference DNS were first validated by comparing them with those of the reference simulation by \cite{kim_turbulence_1987}, for all statistics examined in §\ref{sec:results}, when available. As such, the snapshots can truly be considered to be samples from the coarse-grained natural distribution of the dynamical system (cf. §\ref{sec:natural_distribution}). A second, statistically independent DNS for testing was obtained using the same spin-up procedure and simulation horizon, but with a different random seed for the initial condition.

The machine-learned distribution is trained on a minimal conditional flow unit, which has another resolution and size than the DNS domain.
Here, the fields are coarsened by a factor 2 to a resolution $\Delta x^+ \times \Delta y^+ \times \Delta z^+ = 23.6 \times [0.34-6.2] \times 11.8$, by filtering the field in the periodic directions with a sharp cut-off filter at half the maximal wavenumber and downsampling the field in the wall-normal direction.
To get rid of potential elongated streaks in the average over all snapshots due to the locking of streamwise structures in the periodic domain \citep{munters_shifted_2016}, we also randomly shift the flow field of each snapshot in the periodic $z$-direction, which is consistent with the translational invariance. 
Since the DNS domain is sufficiently larger than the minimal conditional channel flow unit of size $2\upi \delta \times 2 \delta  \times \upi \delta/2 $ (cf. §\ref{sec:minimal-conditional-flow-units}), the statistics within the minimal conditional flow units are effectively homogeneous. 
After splitting the domain into $24$ minimal conditional flow units, the train and test DNS both provide $4344$ velocity field snapshots, which are no longer periodic in the streamwise and spanwise direction. 
As a side effect, energy spectra computed on them will be subject to spectral leakage of the discrete Fourier transform.
Snapshots from the first $40\delta/u_\tau$ after spin-up of the first reference DNS are used to train the machine-learned distribution. Snapshots from the next $5\delta/u_\tau$ are held out as a validation set to detect overfitting during training and select model settings, without being used to update the model parameters. 
All snapshots from the second reference DNS are reserved as an independent test set for evaluating the learned distribution, as discussed in the next section.
Further details on the neural network architecture and training procedure are given in Appendix~\ref{app:ML}.

\section{Applications in synthetic turbulence and turbulent flow reconstruction}\label{sec:results}

Three properties of the machine-learned distribution are examined.
First, the reproduction of essential turbulent statistics is assessed in 
§\ref{sec:results:minimal}, by comparing the statistics of synthetic turbulent channel flow fields on minimal conditional flow units with those of DNS. 
Then, the consistency of conditional sampling is analysed in §\ref{sec:results:large}, for a flow reconstruction problem that allows for a clear probabilistic interpretation of the reconstruction error. In addition, we examine whether the sequential conditional sampling with minimal conditional flow units indeed allows the generation of velocity fields on larger domains.
Lastly, the dynamical invariance of the distribution is studied in §\ref{sec:results:spinup}, by examining the time evolution of the statistics of synthetic velocity fields on a large domain when used as initial conditions for DNS.
The results presented below are obtained by integrating both sets of ordinary differential equations \eqref{eq:trained_generator} and \eqref{eq:conditional_ode} with a fourth-order Runge--Kutta method with only 20 timesteps, which takes about 1 second per field on a Nvidia H100 GPU.

\begin{figure}
	\centering
    \includegraphics[width=0.9\linewidth]{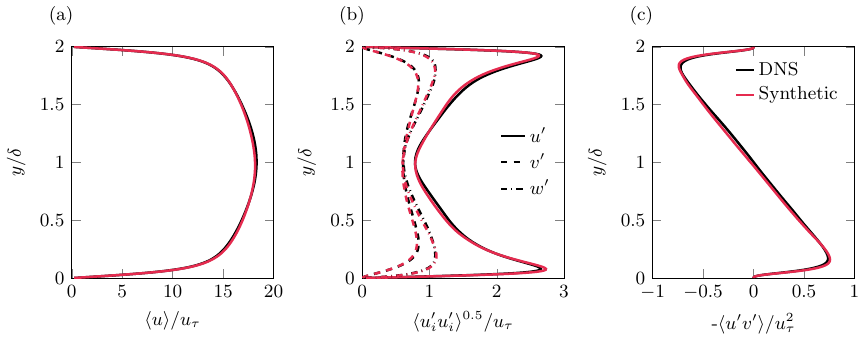}
	\caption{
    Comparison of the one‑point first‑ and second‑order ensemble statistics between the synthetic and the DNS velocity fields: (a) mean velocity profile, (b) root‑mean‑square velocity fluctuations, and (c) Reynolds shear stress.
    }
	\label{fig:minimal-second-order}
\end{figure}

\subsection{Synthetic turbulent channel flow fields}\label{sec:results:minimal}

We start our analysis with figure~\ref{fig:minimal-second-order}, which compares the first- and second-order one-point statistics of an ensemble of 500 synthetic velocity fields on minimal conditional flow units with 500 fields randomly drawn from the unseen DNS test data. Figure~\ref{fig:minimal-second-order}(a) shows that the mean velocity profile is correctly represented, and as a consequence the viscous shear is as well.  Also the turbulent kinetic energy in each of the three velocity components is properly captured as seen in figure~\ref{fig:minimal-second-order}(b). Close to the wall, the root-mean-square velocity fluctuations are highly anisotropic as the variance of the streamwise velocity fluctuations is significantly larger than the others. Also the inhomogeneity in the wall-normal direction is adequately captured. In addition, the generated flow fields display the correct turbulent vertical momentum flux or shear stress profile as shown in figure~\ref{fig:minimal-second-order}(c). Consequently, the synthetic turbulent flow fields convincingly represent the inhomogeneity and anisotropy in the Reynolds stresses. 
{However, these one-point statistics can be reproduced by Gaussian fields as well.}

\begin{figure}
	\centering
	\includegraphics[width=0.8\linewidth]{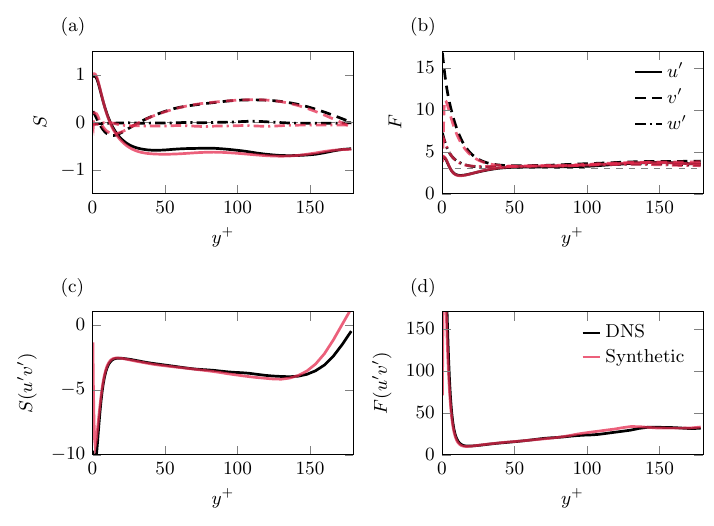}

	\caption{
    Comparison of the one‑point third-, fourth-, sixth- and eighth-order ensemble statistics between the synthetic and the DNS velocity fields: (a,c) skewness $S(\varphi)=\langle \varphi^3 \rangle/\langle \varphi^2 \rangle^{3/2}$, (b,d)  flatness $F(\varphi)=\langle \varphi^4 \rangle/\langle \varphi^2 \rangle^{2}$; (a,b) velocity fluctuations, (c,d) turbulent shear stress (using uncentred moments). 
    The dashed grey lines in (a,b) denote the Gaussian values of $S=0$ and $F=3$ for the third- and fourth-order moments.
    }
	\label{fig:minimal-non-gauss}
\end{figure}

Following \cite{kim_turbulence_1987}, we also examine the higher-order one-point statistics. The skewness $S(\varphi)=\langle \varphi^3\rangle / \langle \varphi^2 \rangle^{3/2}$ of the velocity component fluctuations is compared in figure~\ref{fig:minimal-non-gauss}(a). Close to the wall, the streamwise velocity fluctuation $u'=u-\langle u\rangle$ has large positive skewness while farther from the wall it has negative skewness. This is consistent with the result from quadrant analysis showing a higher contribution to the production of turbulent kinetic energy by sweeps of high-speed fluid towards the wall ($u'>0, v'<0$) close to the wall ($y^+<15$) and by ejections of low-speed fluid outward of the wall ($u'<0, v'>0$) farther from the wall \citep{wallace_quadrant_2016}. The wall-normal velocity fluctuation $v'$ has positive skewness in the viscous sublayer, negative skewness in the buffer layer, and positive skewness in the log-law region. The skewness of the spanwise velocity component fluctuation is zero, because of the spanwise reflection symmetry of the turbulent channel flow dynamics.
The flatness $F(\varphi)=\langle \varphi^4\rangle / \langle \varphi^2 \rangle^{2}$, also known as kurtosis, of the velocity component fluctuations is shown in figure~\ref{fig:minimal-non-gauss}(b). The synthetic turbulent channel flow fields reproduce the highly intermittent behaviour close to the wall, as indicated by the large flatness or heavy tails of their distribution. Very close to the wall, however, the skewness and flatness of the wall-normal velocity in the synthetic fields tend towards the Gaussian values of $S=0$ and $F=3$.
As the velocity fluctuations vanish at the wall and the wall-normal velocity fluctuations are the weakest in terms of root-mean-square magnitude (cf. figure~\ref{fig:minimal-second-order}(b)), this artefact only has a small influence on the discrepancy between true and synthetic fields.
Figure~\ref{fig:minimal-non-gauss}(c) and \ref{fig:minimal-non-gauss}(d) show that the large skewness and flatness of the (uncentred) turbulent shear stress are also replicated, except very close to the wall, similarly to the wall-normal velocity component in figure~\ref{fig:minimal-non-gauss}(b). Nevertheless, the extremely intermittent nature of the Reynolds shear-stress-producing events near the surface is captured. 
Overall, the synthetic turbulent flow fields adequately reproduce the selected one-point statistics reported by \cite{kim_turbulence_1987}, including higher-order moments up to eighth order, reflecting substantial deviations from Gaussianity through significant skewness and intermittency.

\begin{figure}
    \centering
    \includegraphics[width=0.85\linewidth]{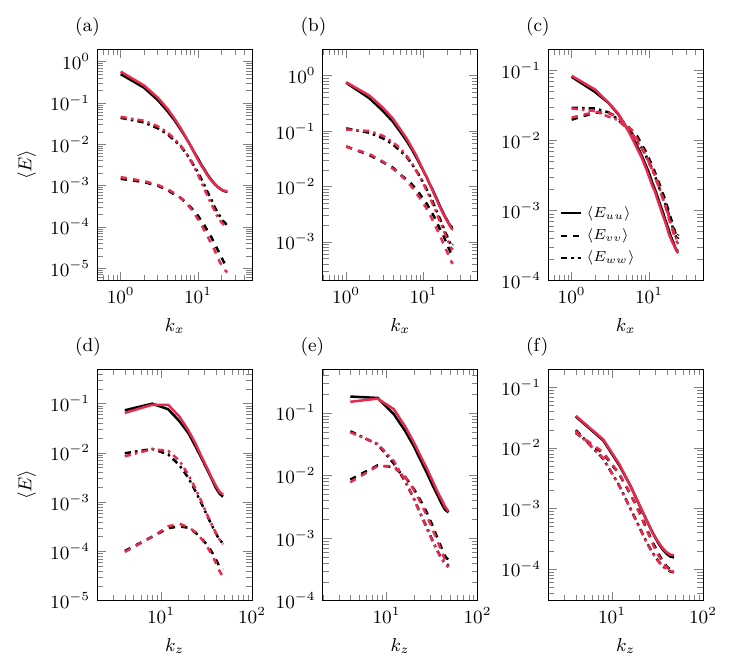}
    \caption{
    Comparison of the one-dimensional energy spectra $\langle E_{uu}\rangle$, $\langle E_{vv}\rangle$, and $\langle E_{ww}\rangle$ between the synthetic (red) and the DNS (black) velocity fields on the minimal conditional channel flow unit of $2\pi \delta \times 2\delta \times \pi\delta/2$ at three wall-normal heights: (a,d) $y^+=5.2$, (b,e) $y^+=29$, (c,f) $y^+=178$.  
    }
    \label{fig:spectra-small}
\end{figure}

\begin{figure}
    \centering
    \includegraphics[width=\linewidth]{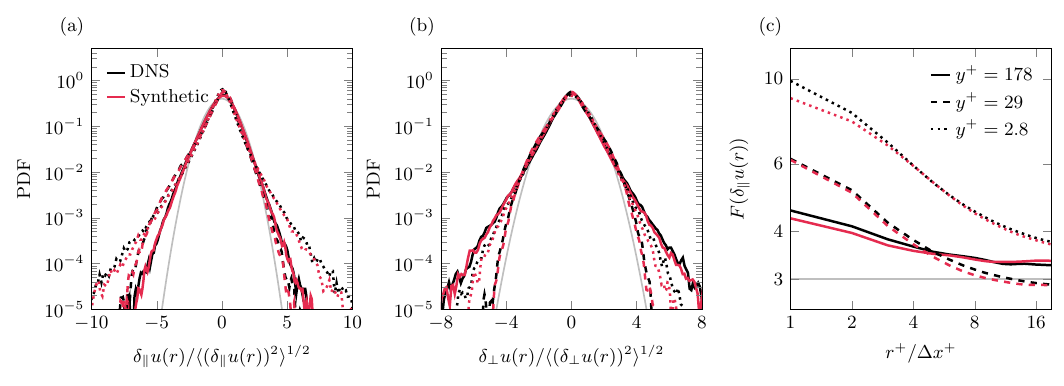}
    \caption{Comparison of the distribution of streamwise velocity increments between the synthetic and the DNS velocity fields: (a) increments in the streamwise direction at $r = \Delta x$ ($\Delta x^+ = 23.6$), (b) increments in the spanwise direction at $r = \Delta z$ ($\Delta z^+ = 11.8$), (c) scale-dependent flatness of the streamwise increments. Grey lines denote Gaussian values and are given for reference. }
    \label{fig:increments}
\end{figure}

The spatial coherency of the synthetic turbulent channel flow fields is examined by comparing their two-point statistics with those of DNS velocity fields out of sample. Figure~\ref{fig:spectra-small} shows the one-dimensional streamwise and spanwise energy spectra of the three velocity components at three wall-normal locations: $y^+\in \{5.2,29,178\}$. Recall that the minimal conditional flow units are not periodic, so the computed spectra are subject to spectral leakage, as seen in the inflated values for $\langle E_{uu} \rangle$ at $y^+=5.2$ and large wavenumbers.\footnote{This artefact becomes more apparent when comparing the reference DNS spectra computed on the small non-periodic domain with those computed on the large periodic domain, shown in figure~\ref{fig:fig11}.} It is seen that the synthetic velocity fields reproduce the spectral energy content of the DNS velocity fields over all scales that can be represented in the considered minimal conditional flow unit. Only the largest spanwise scales of the streamwise velocity component are underrepresented as seen in figure~\ref{fig:spectra-small}(e). By the relation of the energy spectrum to the spatial correlation, the good agreement implies that the spatial coherency up to second order is correctly mimicked.

\begin{figure}
    \centering
    \includegraphics[width=0.9\linewidth]{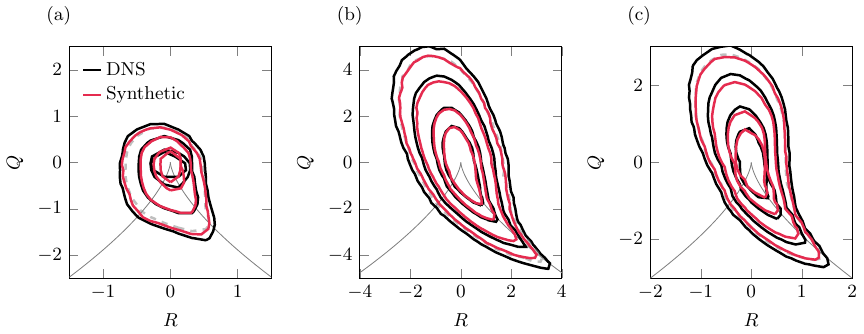}
    \caption{Comparison of the joint distribution of the $Q$ and $R$ invariants of the coarse-grained velocity gradient tensor: (a) $0 < y^+ \leq 5$, (b) $5 < y^+ \leq 35$, (c) $35 < y^+ \leq 180$. 
    The contours are isodensity contours containing  90\%, 95\%, 98\% and 99\% of the probability mass, corresponding to more than 3.8 decades of probability density with respect to the peak. For reference, the Vieillefosse line $Q^3+27R^2/4=0$ (full) and the DNS isodensity contour containing 98.75\% of the probability mass (dashed) are also  shown in grey.}
    \label{fig:qr}
\end{figure}

Reproducing the spectral energy content is also possible with Gaussian random Fourier methods \citep{kraichnan_diffusion_1970, wu_inflow_2017}.
Therefore, we also examine the local structure of the turbulent fluctuations through the distribution of velocity increments in figure~\ref{fig:increments}. In the centre of the channel at $y^+=178$, the distributions of streamwise ($\Delta x^+ = 23.6$) and spanwise increments ($\Delta z^+ = 11.8$) of the streamwise velocity of the synthetic fields displays the same heavy tails as those of DNS. Closer to the wall at $y^+=29$, the synthetic fields also reproduce the large negative skewness of the longitudinal velocity increments, which is related to the interscale energy flux -- cf. Kolmogorov's four-fifths law and the generalised K\'arm\'an--Howarth equation \citep{vassilicos_dissipation_2015}. 
In the viscous sublayer, at $y^+=2.8$, both longitudinal and transversal streamwise velocity increment distributions exhibit strikingly heavy tails, indicating that the near-wall small-scale intermittency is successfully captured by the synthetic fields. Following \cite{benzi_intermittency_1999}, we further assess the flatness of the longitudinal streamwise velocity increments at $y^+\in \{2.8,29,178\}$ as a scale-dependent intermittency measure.
As shown in figure~\ref{fig:increments}(c), the increasingly intermittent character of the increments at smaller scales is reproduced remarkably well by the synthetic fields. 
Overall, the synthetic turbulent velocity fields thus adequately reproduce the large- and small-scale spatial coherency of the velocity fields from DNS, including their distinctly non-Gaussian features.

To compare the incompressibility and local topological features of the coarsened DNS and synthetic velocity fields, we examine the principal invariants $P$, $Q$, and $R$ of their velocity-gradient tensors, following \cite{blackburn_topology_1996}. 
The dimensionless velocity gradient tensor is defined as  $A_{ij}=\partial (u_i/\langle U_c \rangle) /\partial (x_j/\delta)$, where $\langle U_c \rangle$ is the mean centreline velocity magnitude. 
Because the minimal conditional flow unit is non-periodic in the streamwise and spanwise directions, the corresponding derivatives are evaluated using fourth-order finite differences rather than the spectral differentiation used in the original DNS. 
The first invariant, $P=-A_{kk}$, is the negative dimensionless divergence and therefore measures the residual departure from solenoidality \citep{blackburn_topology_1996}.
The normalised divergence measure $\langle\int_\Omega(\nabla\cdot\boldsymbol{u})^2/(3\mathsfbi{S}:\mathsfbi{S})\,\mathrm{d}\Omega\rangle/|\Omega|$ is $5.6\times10^{-3}$ for the coarsened DNS and $5.4\times10^{-3}$ for the synthetic fields, indicating comparably small residual divergence under this discretisation. We remove this residual isotropic component by forming the deviatoric tensor $A_{ij}^{\mathrm{dev}}=A_{ij}-\delta_{ij}A_{kk}/3$, for which $P^{\mathrm{dev}}=-A_{kk}^{\mathrm{dev}}=0$ by construction. The remaining invariants are then calculated as $Q=-A_{ij}^{\mathrm{dev}}A_{ji}^{\mathrm{dev}}/2$ and $R=-A_{ij}^{\mathrm{dev}}A_{jk}^{\mathrm{dev}}A_{ki}^{\mathrm{dev}}/3$, consistent with \citet{blackburn_topology_1996} for $P=0$ and \citet{meneveau_lagrangian_2011}.
Similar to \cite{blackburn_topology_1996}, we aggregate the joint $Q$-$R$ distribution over wall-normal regions without correcting for the non-uniform wall-normal grid spacing, so that points closer to the wall receive greater weight.

Figure~\ref{fig:qr} compares the joint distributions of the $Q$ and $R$ invariants for the synthetic and DNS velocity fields in the viscous sublayer ($0<y^+\leq5$), buffer layer ($5<y^+\leq35$), and logarithmic-to-outer-layer region ($35<y^+\leq180$). Outside the buffer layer, the synthetic fields reproduce the characteristic teardrop shape of the distribution, with good agreement between the isodensity contours enclosing 90 and 95\% of the probability mass. In particular, the contours capture both the $Q>0$, $R<0$ sector, which is associated primarily with vortex stretching, and the extension into the strain-dominated $Q<0$, $R>0$ sector, where strain self-amplification predominates \citep{lozano-duran_multiscale_2016}. The latter extension follows the $Q<0$, $R>0$ branch of the Vieillefosse line, $D=Q^3+27R^2/4=0$, commonly referred to as the Vieillefosse tail \citep{meneveau_lagrangian_2011}. 
The comparable extent of the synthetic and DNS contours on either side of $D=0$ and $R=0$ also indicates similar representation of the four local flow topologies: stable-focus/stretching and unstable-focus/compressing for $D>0$, and stable- and unstable-node/saddle/saddle for $D<0$, with $R<0$ and $R>0$ identifying the stable and unstable cases, respectively \citep{chong_general_1990,blackburn_topology_1996}.
In the buffer layer, differences in the shapes of the 90 and 95\% contours indicate a change in the relative weighting of the most probable physical and topological states, although the close agreement of the 98\% contours shows that the broader range of less frequent states is retained. In all regions, the 99\% DNS contour is slightly wider, indicating a modestly broader distribution of rare velocity-gradient states. Nevertheless, we note that the DNS contour enclosing 98.75\% of the probability mass closely matches the synthetic 99\% contour, corresponding to absolute difference of 0.25 percentage points in the probability mass of the most extreme events.

In summary, the synthetic velocity fields on the minimal conditional flow unit reproduce the first- and second-order one point statistics, as well as selected higher-order statistics up to eighth order \citep{kim_turbulence_1987}, one-dimensional energy spectra, velocity increment distributions and their scale-dependent flatness \citep{benzi_intermittency_1999}, and the statistical distribution of local flow topologies \citep{blackburn_topology_1996}. Moreover, their divergence is comparable to that of the coarsened DNS fields, indicating that the incompressibility constraint is approximately satisfied.

\subsection{Consistency of conditional sampling}\label{sec:results:large}
The second property of interest is the consistency of conditional sampling from the learned distribution. 
This property is relevant, on the one hand, to flow reconstruction problems where a part of the velocity field is observed and another part must be inferred. 
On the other hand, it enables the generation of synthetic turbulent velocity fields on arbitrarily large domains via the sequential conditional sampling procedure illustrated in figure~\ref{fig:generation}(b) and described in §\ref{sec:summary_distribution}. 
To assess this consistency, we focus on three aspects: (1) the spatial evolution of the mean-square-reconstruction error, (2) the realism of the ensemble statistics of the reconstructed fields, and (3) the validity of the local-conditioning approximation in \eqref{eq:local_conditioning}, used in the sequential conditional sampling procedure with minimal conditional flow units. 
The first aspect is examined in a derived flow reconstruction problem in §\ref{sec:flow-rec}, whereas the latter two are evaluated for synthetic turbulent flow fields on a large domain in §\ref{sec:large-dom}.

\begin{figure}
	\centering
    \includegraphics[width=0.75\linewidth]{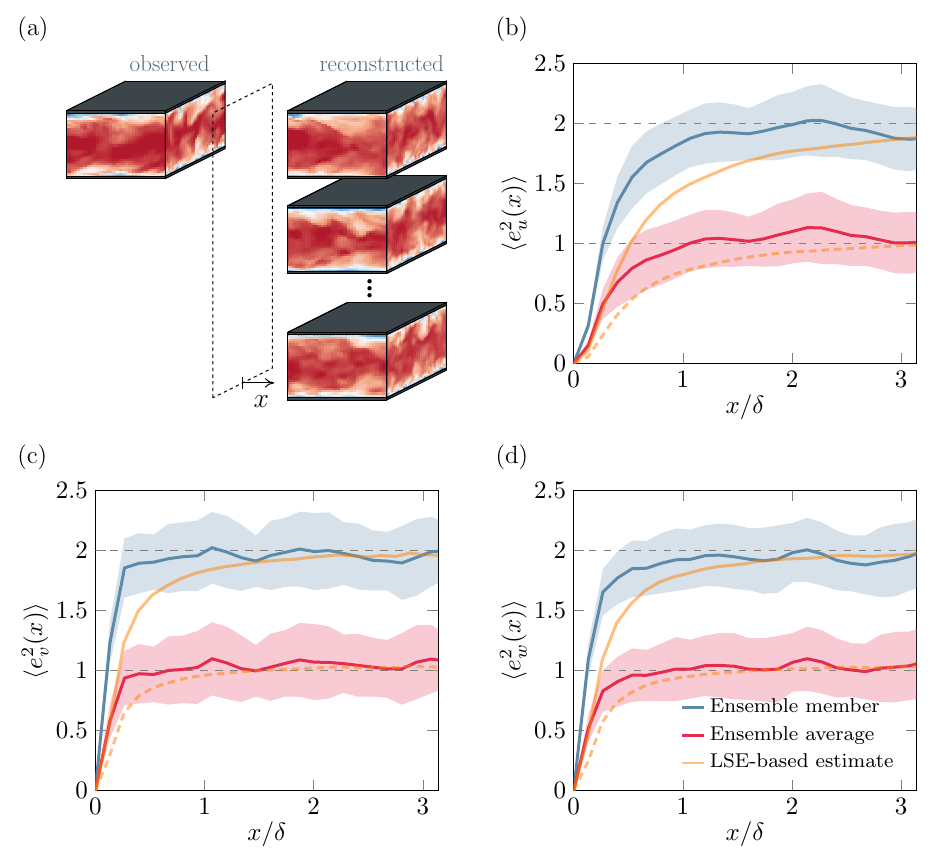}

	\caption{
    Reconstruction error in a flow reconstruction problem: (a) schematic representation of the problem, (b-d) mean-square error on the streamwise, wall-normal, and spanwise velocity component. Shown are the wall-normal averages (mean $\pm$ one standard deviation) for both individual reconstructed fields and the ensemble average, computed from 50 reference fields with $M=20$ reconstructions each. 
    The orange lines denote LSE-based estimates of the reconstruction error for realisations of the true conditional distribution (solid) and for the average of $M$ such realisations (dashed), the latter including the finite-ensemble factor $1+1/M$. } 
	\label{fig:reconstruction}
\end{figure}

\subsubsection{Turbulent flow reconstruction}\label{sec:flow-rec}

Consider the flow reconstruction problem in figure~\ref{fig:reconstruction}(a), in which a downstream half of a minimal conditional channel flow unit is unobserved and should be reconstructed from the observed upstream half.
The true unobserved part of the velocity field necessarily follows the conditional natural distribution, but it is unknown to which particular realisation it corresponds. 
Consequently, an ensemble of fields drawn from this distribution spans the range of dynamically consistent instantaneous fields compatible with the observations, while the associated ensemble variance quantifies the inherent reconstruction uncertainty.

By simulating the ordinary differential equation of \eqref{eq:conditional_ode} with the corresponding mask, we obtain an ensemble of reconstructed velocity fields that exactly equal the reference velocity field in the observed part of the domain, and should remain consistent with the observed part in the unobserved region. 
This consistency is quantified through the mean-square error of every velocity component in the reconstructed part of the domain, normalised by the height-dependent variance of that velocity component and averaged over the channel height. For the streamwise velocity component, this becomes
\begin{equation}\label{eq:normalised-mse}
   \langle e^2_{u}(x)\rangle = \frac{1}{\pi \delta^2} \int_0^{2\delta} \int_0^{\pi\delta /2} \frac{\mathbb{E} \left[ ({u}_\text{rec}(\boldsymbol{x}) - u_\text{ref}(\boldsymbol{x}))^2 \right]}{\left\langle u'(\boldsymbol{x}) u'(\boldsymbol{x})\right\rangle } ~ \mathrm{d} z ~\mathrm{d} y,
\end{equation}
where the expectation in the numerator is taken over the ensemble of reference fields and, for each reference field, over the ensemble of reconstructed fields, whereas the denominator is computed as an ensemble average over the DNS test data.
When computing the integral in the wall-normal direction, we however neglect the error in the region closer than $\Delta y^+=1$ to the wall, because of the vanishing variance.

As a point of reference, we also derive an estimate for the reconstruction error made by realisations of the true conditional distribution, using approximations from linear stochastic estimation \citep{adrian_stochastic_1996,tinney_spectral_2006}. The methodology behind this estimate is presented in appendix~\ref{app:LSE}. 
We consider both the error obtained by reconstructed fields drawn from the true conditional distribution and their $M$-member ensemble mean. 
In principle, the mean-square error of the conditional average represents the lowest attainable reconstruction error. However, because our estimate is based on linear stochastic estimation, the estimated reconstruction error for the conditionally averaged field should be interpreted as the mean-square error of the best linear approximation to the conditional average.

Figures~\ref{fig:reconstruction}(b-d) show the mean-square error obtained by averaging over an ensemble of 20 reconstructed fields for each of 50 reference fields from the unseen test data. The reconstruction error increases gradually with distance into the unobserved region, approaching the marginal variance for the ensemble mean and twice the marginal variance for individual ensemble members. This behaviour is consistent with the probabilistic interpretation of the mean-square reconstruction error: for the conditional mean, it equals the conditional variance (cf.~\citealt{adrian_stochastic_1996}), whereas for fields sampled from the conditional distribution, it equals twice the conditional variance (cf.~§\ref{sec:minimal-conditional-flow-units} and appendix~\ref{app:LSE}). 
Under the LSE approximation, the conditional variance converges to the marginal variance as the quantity of interest decorrelates from the observations.
This convergence occurs more rapidly for the spanwise and wall-normal velocity components than for the streamwise component. Figures~\ref{fig:reconstruction}(b-d) also show the corresponding LSE-based estimate of the reconstruction error, both for the $20$-member ensemble mean and conditional realisations (cf. appendix~\ref{app:LSE}). The averaged reconstruction error of samples from the machine-learned conditional distribution is largely consistent with the LSE-based estimate for samples from the true conditional distribution, although it increases somewhat more rapidly. A similar trend is observed for the error of the ensemble average.
Overall, these results indicate that the conditional sampling procedure provides accurate reconstructions and a satisfactory quantification of the associated uncertainty.

Note that the present definition of a minimal conditional flow unit is specific to conditioning on a single velocity component at the centre of the domain (cf. §\ref{sec:minimal-conditional-flow-units}). Alternative conditioning strategies, involving multiple velocity components and/or spatial locations, may require larger domains before the conditional and unconditional fields become indistinguishable. This is illustrated by the reconstruction experiment in figure~\ref{fig:reconstruction}, where the flow field is conditioned on the upstream half of a minimal conditional flow unit. In this case, the MSE of the streamwise velocity component, $\langle e_u^2(x)\rangle $ in figure~\ref{fig:reconstruction}(b), does not reach twice the marginal variance, indicating that statistically relevant information persists over longer distances than half a conditional minimal flow unit. Nevertheless, as shown below, the current definition proves to be adequate for domain generalisation.

\begin{figure}
	\centering
    \includegraphics[width=0.9\linewidth]{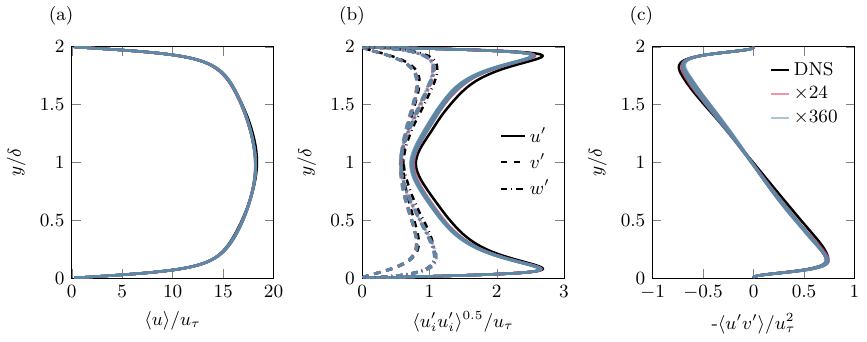}
	\caption{
   Comparison of the one‑point first‑ and second‑order ensemble statistics between the DNS velocity fields and synthetic fields on domains 24 and 360 times larger than the minimal conditional flow unit: (a) mean velocity profile, (b) root‑mean‑square velocity fluctuations, and (c) Reynolds shear stress. For both domains 10 curves are shown, corresponding to five seed-unit locations (centre and four corners) and two tiling orders. Each curve is computed from 20 synthetic fields for $\times 24$ and a single synthetic field for $\times 360$.
    }
	\label{fig:tile}
\end{figure}

\subsubsection{Synthetic turbulence on large domains}\label{sec:large-dom}

The procedure of generating half of a minimal conditional flow unit conditioned on an observed or known other half may be applied sequentially, as depicted in figure~\ref{fig:generation}(b), to generate synthetic velocity fields on arbitrarily large domains. 
With the fast tiling procedure described in §\ref{sec:summary_distribution}, a domain enlargement by a factor $f_x$ in the $x$-direction and a factor $f_z$ in the $z$-direction requires a single integration of the unconditional system \eqref{eq:trained_generator}, $N_\textit{uncond} =1$, and $N_\textit{cond} =4f_x f_z-2f_x-2f_z$ integrations of the conditional system \eqref{eq:conditional_ode}.
For the original DNS domain of $8\pi\delta \times 2\delta \times 3\pi\delta$ ($f_x=4$, $f_z=6$), corresponding to a domain $24$ times larger than the minimal conditional flow unit, this amounts to $77$ integrations, each requiring approximately one second on an Nvidia H100 GPU.
As an even larger-scale example, a domain of $60\pi\delta \times 2\delta \times 6\pi\delta$ ($f_x=30$, $f_z=12$), corresponding to a $360$-fold enlargement of the minimal conditional flow unit and the largest channel-flow DNS domain reported by \citet{lozano-duran_effect_2014}, requires $1357$ integrations and approximately $21$ minutes when performed sequentially on the same hardware.
Figure~\ref{fig:tile} shows that the mean velocity profile and Reynolds stresses are accurately reproduced on domains up to $360$ times larger than the minimal conditional flow unit, although the root-mean-square velocity fluctuations are slightly underpredicted towards the channel centre and top wall. We further verified that these statistics are also insensitive to the choice of seed unit and tiling order. 
The remainder of this section focuses on the original DNS domain, for which a direct comparison with the DNS reference data is possible.

\begin{figure}
    \centering
    \includegraphics[width=\textwidth]{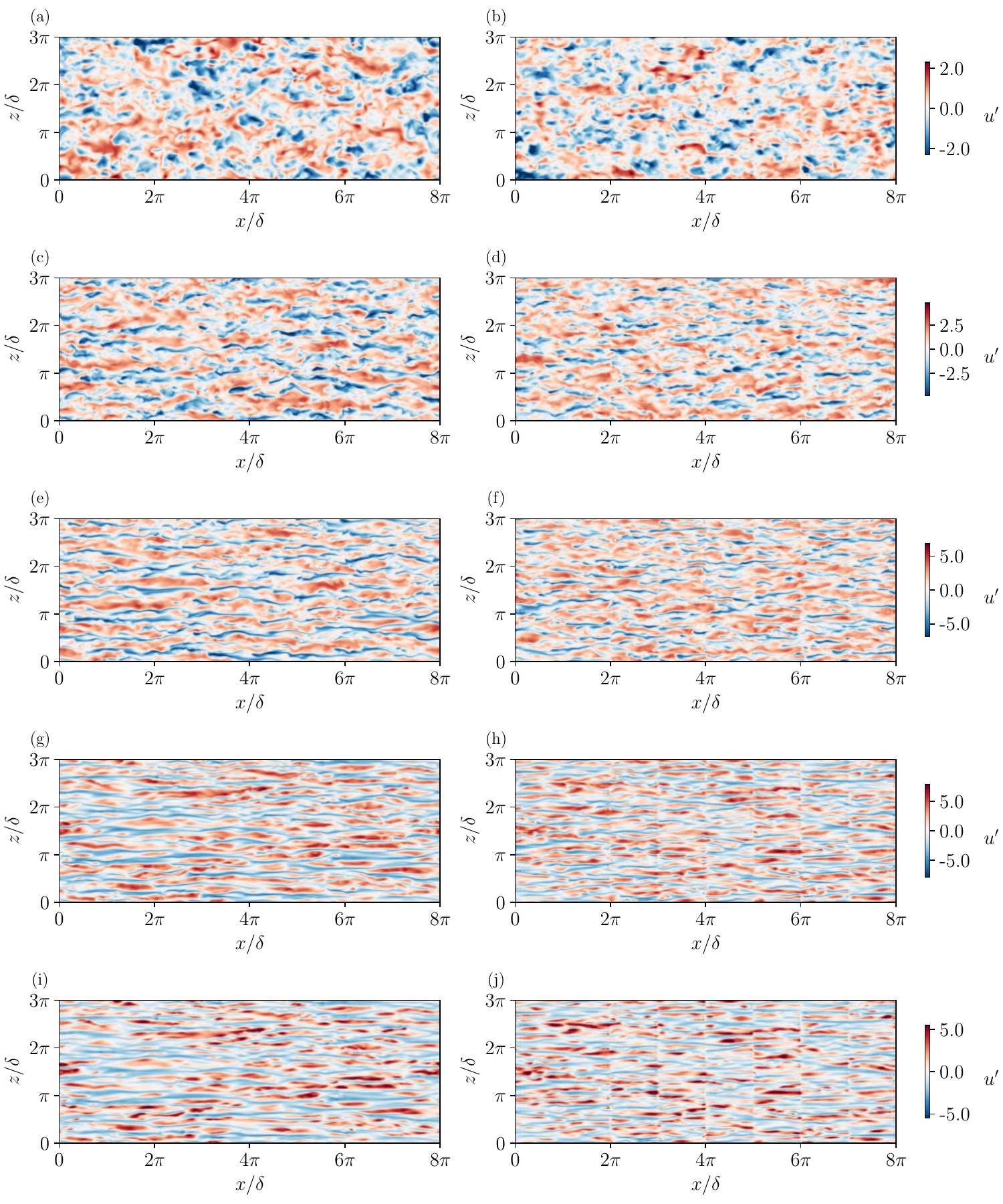} 
	\caption{Comparison of the streamwise velocity fluctuation field between the DNS (left) and the synthetic (right) velocity fields at different wall-normal heights: (a-b) $y^+=178$, (c-d) $y^+=59.5$, (e-f) $y^+=29$, (g-h) $y^+=10$, (i-j) $y^+=5.2$. The synthetic fields employ a synthetic seed unit, located at the origin.}
    \label{fig:uvel}
\end{figure}

\begin{figure}
    \centering
    \includegraphics[width=\textwidth]{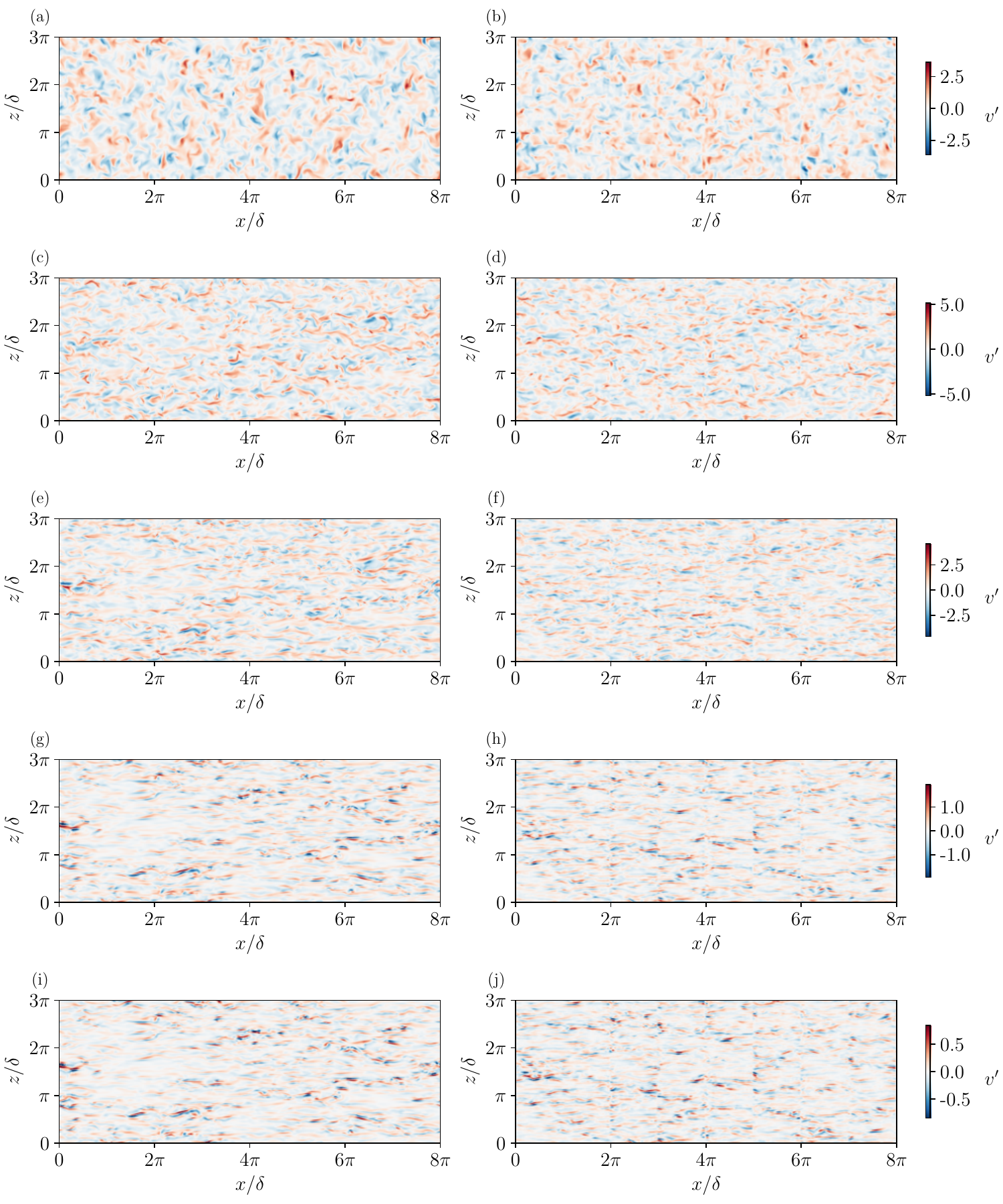} 
	\caption{Comparison of the wall-normal velocity fluctuation field between the DNS (left) and the synthetic (right) velocity fields at different wall-normal heights: (a-b) $y^+=178$, (c-d) $y^+=59.5$, (e-f) $y^+=29$, (g-h) $y^+=10$, (i-j) $y^+=5.2$. The synthetic fields employ a synthetic seed unit, located at the origin.}
    \label{fig:vvel}
\end{figure}

Figure~\ref{fig:uvel} shows the streamwise component of a synthetic velocity field on the original domain size of $8\pi \delta \times 2\delta \times 3 \pi \delta$ to a DNS streamwise velocity field of the unseen test data, at different wall-normal heights.  At the wall-normal heights $y^+=59.5$ and $y^+=29$, as shown in figure~\ref{fig:uvel}(c-f), the elongated structures are qualitatively very similar to those in the DNS, even though they surpass the streamwise length of the minimal conditional flow unit. At the centre of the channel ($y^+=178$), as shown in figure~\ref{fig:uvel}(a-b), the largest spanwise scales appear to be slightly underrepresented. Closer to the wall at $y^+=10$ and $y^+=5.2$, as  shown in figure~\ref{fig:uvel}(g-j), the very large scale motions are generally reproduced, although they are somewhat shortened or cut in the synthetic fields. These cuts at $x/\delta = i\pi$ with $i=2,\ldots, 7$ originate from the boundary of the observed and unobserved region, over which the velocity-increment distributions have heavier tails than in DNS at some, but not all, wall-normal heights (not shown).
The wall-normal velocity component of the synthetic velocity field is shown in figure~\ref{fig:vvel}, again next to a field from the DNS test data. Overall, the shape of the turbulent structures shows strong qualitative resemblance to those of the DNS. Similarly to the streamwise velocity component, some structures may be noticed to be cut through and patched to others at the boundaries of the known and unknown region in the conditional sampling steps. Close to the wall, the highly intermittent behaviour of the wall-normal velocity is also reproduced with strong local bursts amidst more moderate elongated structures. Consequently, sequential conditional sampling with minimal conditional flow units allows generating synthetic turbulent velocity fields that share at least the qualitative intricate features of DNS fields on arbitrarily large domains.

\begin{figure}
	\centering
    \includegraphics[width=0.85\linewidth]{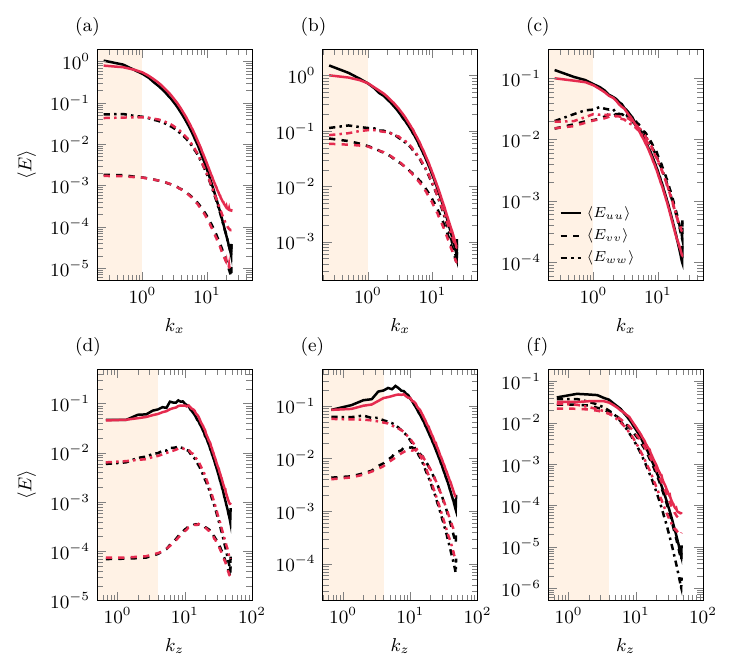} 
    \caption{
    Comparison of the one-dimensional energy spectra $\langle E_{uu}\rangle$, $\langle E_{vv}\rangle$, and $\langle E_{ww}\rangle$ between the synthetic (red) and the DNS (black) velocity fields on a domain of $8\pi \delta \times 2\delta \times 3\pi\delta$ at three wall-normal heights: (a,d) $y^+=5.2$, (b,e) $y^+=29$, (c,f) $y^+=178$. The scales larger than those represented on the minimal conditional flow unit are shaded. 
    }
    \label{fig:fig11}
\end{figure}

The spatial coherency of synthetic turbulent velocity fields generated on the original domain is validated by comparing their spectral energy content with that of the DNS fields. 
Figure~\ref{fig:fig11} shows the one-dimensional streamwise and spanwise energy spectra of the three velocity components at three wall-normal locations: $y^+\in \{5.2,29,178\}$ for an ensemble of 200 synthetic fields and all the test DNS fields.
It is seen that the energy content at the small scales is still well-reproduced for the three velocity components, except close to the wall for the streamwise spectrum of $u$ and $w$ in figure~\ref{fig:fig11}(a), and in the middle of the channel for the spanwise spectrum of the three velocity components in figure~\ref{fig:fig11}(f). This is in agreement with the observed structures that are cut-through and patched together in figures \ref{fig:uvel} and \ref{fig:vvel}. It must be noted, however, that the synthetic turbulent flow fields do not satisfy periodic boundary conditions and are thus also subject to spectral leakage. The same issues of inflated energy content were indeed observed for the DNS fields on minimal conditional flow units in figure~\ref{fig:spectra-small}(a,f). 
Even the energy content at scales larger than those represented in the minimal conditional flow unit (shaded in figure~\ref{fig:fig11}), is adequately reproduced by the synthetic velocity fields. The longest streamwise and spanwise structures of the streamwise velocity are, however, slightly underrepresented at all wall-normal distances. 
The good agreement for the spectra of the other velocity components is to be expected, since the fields are essentially uncorrelated at these distances. This is reflected by the flat one-dimensional spectra at low wavenumbers, which are also obtained for Dirac delta correlations.
This does not imply that uncorrelated units can simply be patched together, as this would break local consistency across their boundaries, which is largely preserved here (cf. figure~\ref{fig:reconstruction} and figure~\ref{fig:fig11} at small scales). Conversely, periodically repeating a single unit would leave no energy at scales larger than the minimal unit.
Overall, the spatial coherency of the synthetic velocity fields is thus well-preserved for the scales represented within the minimal conditional flow unit and generalised up to a domain that is 24 times as large. Consequently, the partial conditioning with minimal conditional flow units serves as an adequate approximation to full conditioning in this setting.

\begin{figure}
	\centering
    \includegraphics[width=0.85\linewidth]{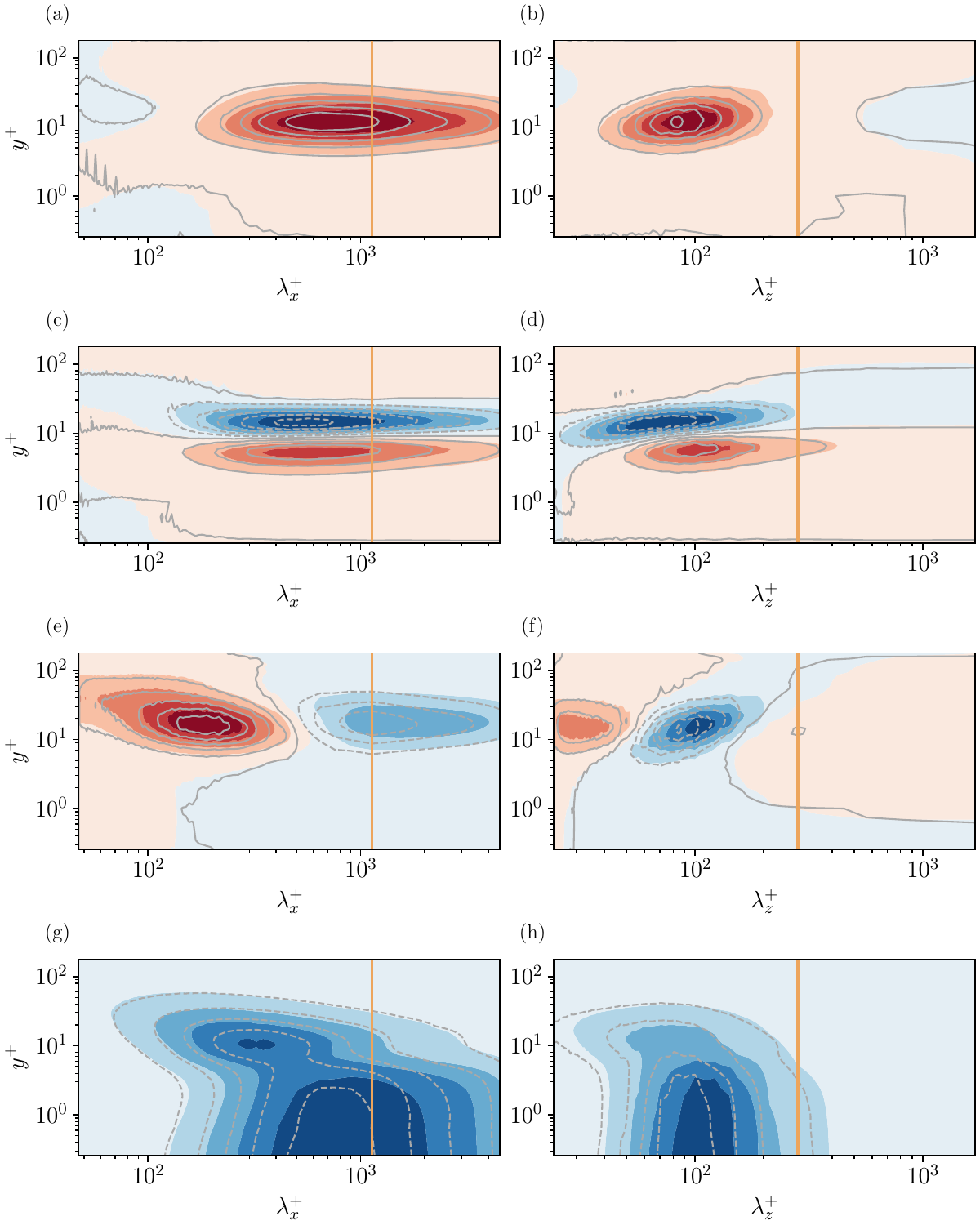} 
    \caption{
    Comparison of the premultiplied one-dimensional cospectra obtained from the synthetic and DNS velocity fields on a domain of size $8\pi \delta \times 2\delta \times 3\pi\delta$: (a,b) production, (c,d) wall-normal transport, (e,f) interscale transport, and (g,h) dissipation of turbulent kinetic energy. In each panel, eleven symmetric, uniformly spaced contour levels are shown. The DNS results are displayed as filled contours (red positive, blue negative), whereas the synthetic results are overlaid using the same contour levels (grey contours; solid positive, dashed negative). The orange vertical lines denote the largest scales that can be represented on the minimal conditional flow unit.
    }
    \label{fig:figX}
\end{figure}

Although the energy spectra demonstrate that the energy content is correctly distributed across scales, they do not by themselves establish whether energy is transported correctly between scales or through the domain. To examine these processes, we consider the scale-dependent production, dissipation, wall-normal transport and interscale transport of turbulent kinetic energy following \citet{mizuno_spectra_2016}, to which the reader is referred for the definitions of these terms. The derivatives in the streamwise and spanwise directions are evaluated spectrally for both the DNS and synthetic fields, resulting in boundary effects and spectral leakage for the latter. Figure~\ref{fig:figX} compares the streamwise and spanwise spectra of the four budget terms as functions of wall-normal distance.
The production spectra shown in panels (a,b) are in close agreement, including the dominant buffer-layer peak at $\lambda_x^+ \sim 500-1000$ and $\lambda_z^+\sim 100$, as well as the regions of negative production near the wall and around $y^+\approx 20$, which have also been observed at higher Reynolds numbers \citep{lee_spectral_2019}.
Notably, these features are accurately recovered even at scales that exceed the extent of the minimal conditional flow unit.
The wall-normal transport spectra in panels (c,d) show that energy is primarily transferred towards the wall and deposited near $y^+\approx 5$. 
Panels (e,f) demonstrate that energy is transferred predominantly from larger to smaller scales, although the spanwise spectra additionally exhibit a region of transfer towards larger scales. Finally, the dissipation spectra in panels (g,h) show that dissipation is concentrated close to the wall and, farther from the wall, occurs predominantly at the smaller scales to which energy is transferred through the interscale transport.

Overall, the synthetic fields accurately reproduce principal mechanisms by which turbulent kinetic energy is produced, transported and dissipated. Moreover, this agreement is maintained well beyond the scales directly represented in the minimal conditional flow unit, demonstrating successful generalisation of the energy-transfer dynamics to larger domains.

\subsection{Dynamical invariance of the machine-learned probability distribution}\label{sec:results:spinup}
We showed that the synthesised and reconstructed velocity fields sampled from the machine-learned distribution are turbulent-looking up to the considered ensemble statistics, but evaluating all ensemble statistics is impracticable. Therefore, we now examine whether the learned distribution is invariant under the turbulent dynamics. Apart from being a defining property of the natural distribution, it is also crucial for synthetic turbulence applications. If the learned distribution is dynamically invariant, initial conditions sampled from the natural distribution remain on the attractor without exciting large transients.
Hence, a practical approach to test this last property of interest is evaluating the stationarity of the ensemble statistics of the previously generated synthetic turbulent fields (cf. figures~\ref{fig:uvel}-\ref{fig:fig11}), when used as initial conditions for DNS.

\begin{figure}
	\centering
	\includegraphics[width=\linewidth]{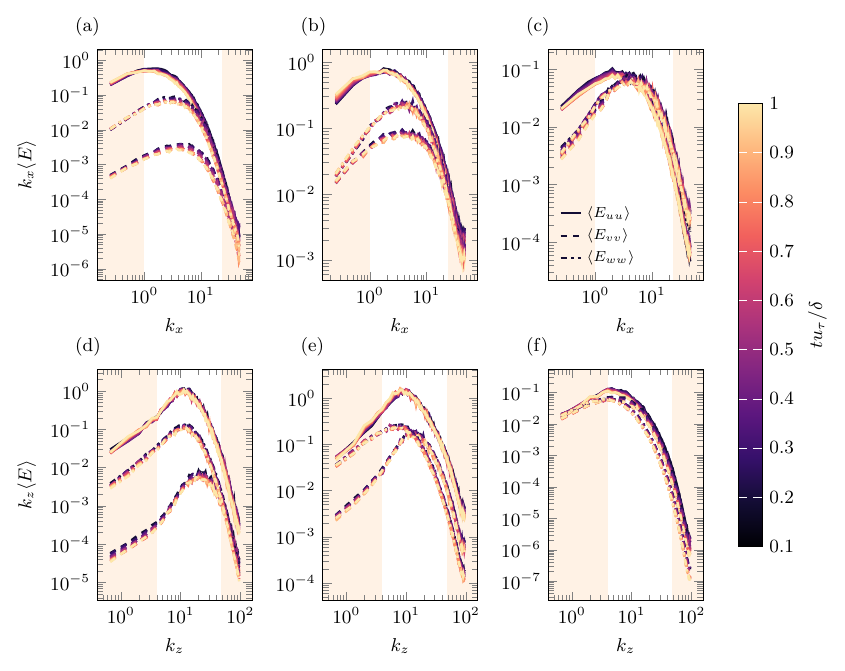}

	\caption{
    Time evolution of the premultiplied one-dimensional energy spectra averaged over an ensemble of 10 trajectories starting from synthetic initial conditions on a domain of $8\pi \delta \times 2\delta \times 3\pi\delta$. The spectra are shown at three wall-normal heights: (a,d) $y^+=5.2$, (b,e) $y^+=29$, (c,f) $y^+=178$. The line colour indicates the time of evaluation, with intervals of  $\Delta t u_\tau/\delta = 0.1$. The scales larger and smaller than those represented on the minimal conditional flow unit are shaded. 
    }
	\label{fig:spectra-time}
\end{figure}

The resolution of the synthetic turbulent velocity fields is fine enough to represent all dynamically relevant scales up to the dissipation range, but not necessarily equal to the DNS resolution, which needs to be finer so as to keep discretisation errors sufficiently small. To obtain initial conditions on the original DNS resolution, we must, in our case, upsample the synthetic turbulent velocity fields by a factor two. 
In the streamwise and spanwise direction, the velocity field is interpolated spectrally. As a result, the smallest scales of the upsampled initial condition initially contain no energy.  
An initial transient is therefore required to populate the smallest scales in the dissipation range. This transient is expected to evolve on the Kolmogorov time scale $t_\nu$, which varies from $0.013\delta/u_\tau$ at the walls to $0.076\delta/u_\tau$ at the centre of the channel, based on the DNS data of \cite{kim_turbulence_1987}. 
Since the synthetic velocity fields are not necessarily divergence-free in the considered discretisation, we also apply a pressure projection step before integrating the Navier--Stokes dynamics.
Before the projection step, the large synthetic fields exhibit an average normalised divergence of $( \int (\nabla \cdot \boldsymbol{u})^2 /(3\mathsfbi{S}:\mathsfbi{S})~\mathrm{d}\Omega   ) /|\Omega|\approx 10^{-2}$, 
computed with spectral differentiation in the streamwise and spanwise directions.
Given that $(\nabla \cdot \boldsymbol{u})^2/(3~\mathsfbi{S}:\mathsfbi{S}) \in [0,1]$ by the Cauchy–Schwarz inequality, this value is small, indicating that the synthetic fields are already close to divergence-free. Accordingly, the pressure projection removes 1.3\% of the turbulent kinetic energy on average. 
Consequently, the implicitly learned approximate incompressibility constraint persists on the large domain.
To assess whether the ensemble statistics reach stationarity after the brief transient, we examine their temporal evolution using DNS snapshots recorded at intervals of $\Delta t u_\tau / \delta = 0.1$ for an ensemble of 10 synthetic turbulent velocity fields.

\begin{figure}
	\centering
	\includegraphics[width=\linewidth]{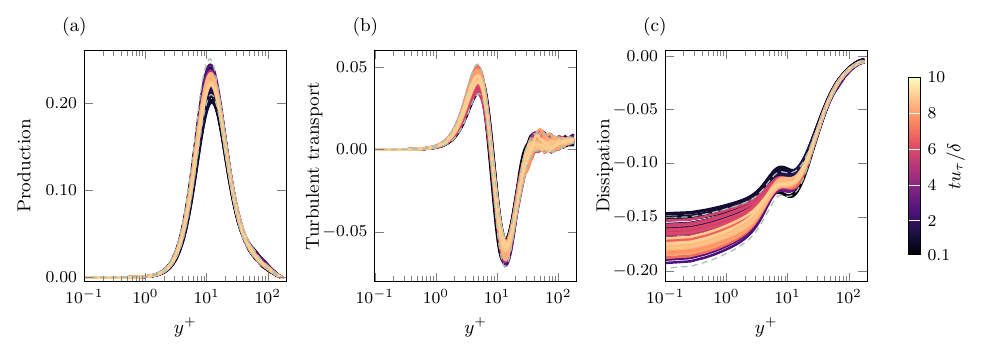}
	\caption{
    Time evolution of turbulent kinetic energy budget terms from a DNS on a domain of $8\pi \delta \times 2\delta \times 3\pi\delta$ initialised with a synthetic turbulent velocity field: (a) production, (b) turbulent transport, (c) dissipation. The line colour indicates the time of evaluation, with intervals of  $\Delta t u_\tau/\delta = 0.1$. The grey dashed lines indicate the 0.5th and 99.5th percentiles from the reference DNS trajectory.  
    }
	\label{fig:prod-diss}
\end{figure}

Rather than attempting an exhaustive statistical characterisation,
we focus on the temporal evolution of the Turbulent Kinetic Energy (TKE), whose spectra, production and dissipation jointly capture the essential features of the energy dynamics. 
Figure~\ref{fig:spectra-time} shows the premultiplied one-dimensional streamwise and spanwise energy spectra at different wall-normal heights, where the premultiplication enables an energy-density interpretation on the logarithmic wavenumber axes.
It is seen that the smallest scales are indeed regenerated after only $0.1\delta/u_\tau \sim \mathcal{O}(t_\nu)$. 
Beyond this point, no significant transient behaviour is observed in the spectral energy content.
Similarly, we find that the near-wall skewness and flatness of the wall-normal velocity component (cf. figure~\ref{fig:minimal-non-gauss}) recover their DNS profiles within this adaptation period (not shown).
In figure~\ref{fig:prod-diss}, we further verify the stationarity of the production and dissipation of TKE, which govern the net evolution of the total TKE, over a longer trajectory spanning $10\delta/u_\tau$. 
Their temporal variations are consistent with those observed in DNS snapshots from the reference trajectory for testing, whose 0.5th and 99.5th percentile are shown as grey dashed lines.
Figure~\ref{fig:prod-diss}(b)  additionally shows that the nonlinear turbulent transport of TKE also remains stationary. 
Since this term depends on third-order moments of the velocity fluctuations, it vanishes identically in Gaussian synthetic fields. This again highlights the importance of the non-Gaussian features of the natural distribution imposed by the turbulent dynamics. 
Overall, these results suggest that the machine-learned distribution exhibits the dynamical invariance property central to the definition of the natural distribution.

\section{Conclusion} \label{sec:conclusion}
Motivated by applications in synthetic turbulence generation and turbulent flow reconstruction, we examine to what extent a machine-learned distribution can approximate the physical invariant distribution of a turbulent channel flow at a friction Reynolds number of $\Rey_\tau=180$.
We study three properties of the approximation: dynamical invariance \citep{hopf_statistical_1952}, ensemble statistics that are representative for observable trajectories of the dynamical system \citep{ruelle_what_1978}, and consistency of conditional sampling, which is particularly relevant to flow reconstruction.
Since training (or conditioning) the distribution on arbitrarily large domains is impracticable, we represent the distribution on a minimal conditional flow unit, and show that this still enables generating velocity fields on larger domains. 
We define the minimal conditional flow unit as the smallest domain in the homogeneous direction(s) outside which the conditional velocity field, conditioned on a velocity component at the domain centre, is indistinguishable from corresponding unconditional fields in terms of mean-square discrepancy from other conditional realisations.
In this way, the minimal conditional flow unit spans the extent over which conditional flow fields, including the conditional eddies from linear stochastic estimation \citep{adrian_stochastic_1996}, are distinguishable from unconditional fields in terms of mean-square discrepancy.

Our methodology is based on training a flow-based generative model \citep{ lipman_flow_2023} to synthesise and reconstruct spatially three-dimensional turbulent channel flow velocity fields. 
Similar to the Navier--Stokes dynamics during spin-up, the flow-based generative model is intended to deterministically map an initial condition to a state on the turbulent attractor. However, it does so by defining an auxiliary dynamical system for the velocity field that is trained on straight trajectories in phase space. 
Random fields that initially follow a standard Gaussian distribution can thereby be transformed into synthetic turbulent channel flow fields by integrating the auxiliary dynamical system, with only 20 steps of a fourth-order Runge-Kutta method in our test cases.
In flow reconstruction problems in which only a part of the velocity field is observed, the unobserved part of the field necessarily follows the conditional natural distribution. 
Hence, an ensemble of realistic reconstructions can be obtained by conditionally sampling the natural distribution, where the ensemble variance quantifies the remaining reconstruction uncertainty.
We show that the conditional learned distribution can be constructed for arbitrary partitions of observed and unobserved variables without retraining a neural network.
This is achieved by deriving a modified dynamical system that approximately minimises a guided conditional flow‑matching loss.
As a result, the conditional natural distribution can be sampled approximately by integrating this modified dynamical system, using the same number of integration steps as for unconditional generation.

The machine-learned distribution displays the three considered properties of the natural distribution, relevant to synthetic turbulence generation and turbulent flow reconstruction. 
First, synthetic turbulent velocity fields sampled from the distribution reproduce the mean velocity, Reynolds stresses, and one-dimensional energy spectra, similarly to classical Gaussian random Fourier methods originating from the work of \cite{kraichnan_diffusion_1970}. Importantly, they additionally capture distinct non-Gaussian features, including the skewness and flatness of the velocity fluctuations and turbulent shear stress, velocity-increment distributions and their scale-dependent intermittency, as well as the distribution of the $Q$-$R$ invariants of the velocity gradient tensor. Moreover, their divergence is comparable to that of coarsened DNS fields, indicating that the incompressibility constraint is approximately satisfied.
Second, conditional sampling is consistent, as the mean-square reconstruction error obtained from an ensemble of reconstructed fields largely agrees with an estimated reconstruction error based on linear stochastic estimation \citep{adrian_stochastic_1996, tinney_spectral_2006}. 
By means of sequential conditional sampling, we also generate synthetic velocity fields on the large original DNS domain, corresponding to a domain enlargement by a factor of 24 relative to the minimal conditional flow unit. 
These fields share the qualitative features of DNS velocity fields and adequately replicate both their energy spectra and scale-dependent wall-normal and interscale transport of turbulent kinetic energy, supporting the representation of the learned distribution on a minimal conditional flow unit. Furthermore, the domain extension procedure remains stable on domains up to 360 times larger than the minimal conditional flow unit.
Third, the synthetic fields on the large domain show no large transient in the examined statistics when used as initial conditions for DNS, suggesting that they remain close to the turbulent attractor.
After an adaptation period of the order of the Kolmogorov timescale, the ensemble-averaged energy spectra remain approximately stationary. Moreover, the turbulent production and dissipation, as well as the nonlinear transport of turbulent kinetic energy, remain statistically stationary over a simulation spanning $10\delta /u_\tau$.
Together, these results show that the machine-learned distribution reproduces key statistical and dynamical features of turbulent channel flow at $\Rey_\tau=180$, is consistent under conditional sampling, and remains close to the turbulent attractor under Navier--Stokes evolution. 
It therefore provides a physically meaningful approximation to the natural distribution of the turbulent dynamical system.

Probabilistic machine-learning models, and flow-based generative models in particular, provide promising computational methods for improving synthetic turbulence generators and flow-reconstruction algorithms, because they can represent intricate features of the full natural distribution. 
The present study focused on the distribution in the phase space of turbulent channel flow at a fixed, relatively low Reynolds number. 
Extending the approach to higher Reynolds numbers may require architectural improvements to handle the rapid increase in the dimensionality of the discretised fields. 
In this regime, it may also be advantageous to impose incompressibility explicitly, rather than relying on the model to learn this constraint implicitly. 
Parametrising the dependence of the distribution on the Reynolds number is another useful direction, although this may require explicit treatment of inner--outer scaling. Further developments are also needed for more practical flow configurations. In particular, unstructured grids are more common in complex geometries than the structured grid used here, and may require a different neural-network backbone. Although the present work defined the conditional distribution of one part of the velocity field given another, conditioning on derived quantities, such as wall shear stress or pressure, may also be valuable. Lastly, the use of large generated fields in combination with Taylor's frozen-turbulence hypothesis may also be of interest for synthetic inflow generation.

\begin{bmhead}[Funding.]
The authors acknowledge support from Research Foundation - Flanders (FWO, grant no. 11A6S26N) and from KU Leuven (project no. 3E230463).
The computational resources and services used in this work were provided by the VSC (Flemish Supercomputer Center), funded by the Research Foundation - Flanders (FWO) and the Flemish Government.
\end{bmhead}

\begin{bmhead}[Declaration of interests.]
The authors report no conflict of interest.
\end{bmhead}

\begin{bmhead}[Author ORCIDs.]
F. Aerts, \href{https://orcid.org/0009-0001-4853-2714}{https://orcid.org/0009-0001-4853-2714}; D. Nuyens, \href{https://orcid.org/0000-0002-4555-2314}{https://orcid.org/0000-0002-4555-2314}; J. Meyers, \href{https://orcid.org/0000-0002-2828-4397}{https://orcid.org/0000-0002-2828-4397}

\end{bmhead}

\begin{appen}

\section{Network architecture and training}\label{app:ML}
The flow‑based generative model is defined by the generator $\boldsymbol{f}_{\!\boldsymbol{\vartheta}}(\boldsymbol{u}, \xi)$, parametrised by a neural network. 
As neural network architecture, we adopt the U‑Net design of \cite{dhariwal_diffusion_2021}, with the selected hyperparameters listed in table~\ref{tab:unet_hyperparameters}. 
The network is trained by minimising the conditional flow-matching loss in \eqref{eq:CFM}, using snapshots from the first 40$\delta/u_\tau$ after spin-up of the first reference DNS as samples of the target distribution. 
Training is performed on a single Nvidia H100 GPU using the R-Adam optimiser \citep{liu_variance_2021}, with an initial learning rate of 1e-4, a batch size of 12, and a total of 500 epochs, amounting to a training time of approximately 18 hours. 
During training, the conditional flow-matching loss decreases steadily on both the training set and the validation set, which consists of snapshots from the subsequent 5$\delta/u_\tau$ of the first reference DNS. 
At the end of each epoch, we generate an ensemble of 48 velocity fields by integrating the current generator from $\xi=0$ to $\xi=1$, using a fourth-order Runge-Kutta scheme with 20 integration steps. 
We then evaluate the mean-square error of the ensemble-averaged velocity field and Reynolds stresses relative to the corresponding statistics computed from 48 velocity fields drawn from the validation set.
The model checkpoint with the lowest validation error for these one-point first- and second-order statistics is selected as the trained flow-based generative model.
Snapshots from the second independent DNS, which are not used during training or validation, are reserved for examining the properties of the machine-learned distribution in~§\ref{sec:results}.

\begin{table}
\centering
\begin{tabular}{llc}
{Hyperparameter} & {Description} & {Values Used} \\[3pt]

Image resolution & Field resolution ($n_x \times n_y \times n_z $) & $ 48\times128\times24$ \\

Input/Output channels & State dimensionality (velocity) & $3$  \\

Model channels & Internal state dimensionality & $64$ \\

Number of resolutions & U-Net levels & 4 \\

Channel multipliers & Multipliers of the base channel count per level & $(1,2,3,4)$\\

Conv resampling & Learned convolutions for up/downsampling & Enabled \\

Residual blocks per level & Number of residual blocks per downsample & $1$ \\

Attention resolutions & Downsample rates with 
        attention blocks & Only at the bottleneck \\

Attention heads & Number of attention heads & $4$  \\

Dropout & Dropout probability in residual blocks & $0.1$ 
\end{tabular}
\caption{Hyperparameters used in the U-Net architecture of \cite{dhariwal_diffusion_2021}.}
\label{tab:unet_hyperparameters}
\end{table}

\section{Estimated reconstruction error based on linear stochastic estimation}\label{app:LSE}
In what follows, we estimate the mean-square error (MSE) between the true unobserved part of the field and a stochastic estimate for it. 
We first relate this MSE to the conditional average and conditional variance, through a multivariate generalisation of \eqref{eq:cond_var}.
Then we describe a procedure of estimating this MSE from the DNS snapshots restricted to a minimal conditional flow unit.
As in §\ref{sec:conditional_sampling}, we denote by $\boldsymbol{{u}_o}=\mathsfbi{M}\boldsymbol{{u}}$  the observed part of the discretised velocity field $\boldsymbol{u} \in \mathbb{R}^{n}$, where $\mathsfbi{M} \in \mathbb{R}^{ m \times n}$ is a mask matrix containing $m$ distinct rows of the identity matrix. The unobserved part of the velocity field is denoted by $\boldsymbol{{u}_u}=\mathsfbi{M}^\perp \boldsymbol{{u}}$, where the complementary mask $\mathsfbi{M}^\perp \in \mathbb{R}^{(n- m) \times n}$ contains the remaining $n-m$ rows of the identity matrix.

\subsection{Expression for the mean-square reconstruction error}
The true unobserved part of the velocity field $\boldsymbol{u}_u^\mathrm{true} $ necessarily follows the conditional distribution $ \pi(\boldsymbol{u}_u | \boldsymbol{u}_o)$ with mean $\langle \boldsymbol{u}_{u} | \boldsymbol{u}_{o}\rangle$ and covariance $\operatorname{Cov}(\boldsymbol{u}_{u} | \boldsymbol{u}_{o})$.
Let the reconstruction $\boldsymbol{u}_u^\mathrm{rec}$ be drawn independently from a distribution with mean $\langle \boldsymbol{u}_u^\mathrm{rec} | \boldsymbol{u}_{o} \rangle$ and covariance $\operatorname{Cov}(\boldsymbol{u}_{u}^\mathrm{rec} | \boldsymbol{u}_{o})$. 
Because the true realisation $\boldsymbol{u}_u^\mathrm{true}$ is unknown, we examine the expected squared error by averaging over all possible $\boldsymbol{u}_u^\mathrm{true}$. Likewise, when the reconstruction is stochastic, we average over all possible reconstructions $\boldsymbol{u}_u^\mathrm{rec}$.  
Because the reconstruction error depends on the observed state, we also average over $\boldsymbol{u}_o $, so that the resulting error quantifies ensemble-averaged performance rather than performance conditional on a particular observation.
The resulting MSE is given by
\begin{equation}
     \left\langle \| \boldsymbol{u}_u^\mathrm{rec} - \boldsymbol{u}_u^{\mathrm{true}}  \|^2  \right\rangle =  \mathbb{E}_{\boldsymbol{u}_{o}} \left[ \operatorname{Tr} \left( \operatorname{Cov}(\boldsymbol{u}_u^\mathrm{rec} | \boldsymbol{u}_o) \right) +  \operatorname{Tr} \left( \operatorname{Cov}(\boldsymbol{u}_u | \boldsymbol{u}_o) \right) +  \| \langle  \boldsymbol{u}_u^\mathrm{rec} |  \boldsymbol{u}_o \rangle - \langle \boldsymbol{u}_u | \boldsymbol{u}_o \rangle \|^2 \right],
\end{equation}
which follows from the variance identity.  
For a particular observation $\boldsymbol{u}_o$, the lowest mean-square error of all deterministic estimators is achieved by the conditional average \citep{adrian_stochastic_1996}. Indeed,
\[
       0 = \frac{\partial}{\partial  \boldsymbol{u}_{u}^{\mathrm{rec}}} \langle   \|  \boldsymbol{u}_{u}^{\mathrm{rec}} -\boldsymbol{u}_{u}^\mathrm{true} \|^2 | \boldsymbol{u}_o \rangle
      = -  2 ( \langle \boldsymbol{u}_{u} | \boldsymbol{u}_o \rangle -  \boldsymbol{u}_{u}^{\mathrm{rec}})  
      \Leftrightarrow 
       \boldsymbol{u}_{u}^{\mathrm{rec}} = \langle \boldsymbol{u}_{u} | \boldsymbol{u}_o \rangle.
\]
However, the conditional average is not necessarily representative of instantaneous turbulent structures. Like the true unobserved part of the velocity field, realistic conditional fields should follow the conditional natural distribution. For an independent realisation of the conditional velocity field $ \boldsymbol{u}_{u}^{c}\sim \pi(\boldsymbol{u}_u | \boldsymbol{u}_o)$, the mean-square error becomes
\begin{equation}\label{eq:mse_conditional_rec}
        \left\langle \| \boldsymbol{u}_{u}^{c} - \boldsymbol{u}_u^{\mathrm{true}}  \|^2  \right\rangle =  2  \operatorname{Tr} \left( \mathbb{E}_{\boldsymbol{u}_{o}} \left[ \operatorname{Cov}(\boldsymbol{u}_u | \boldsymbol{u}_o) \right] \right) .
\end{equation}
Similarly, the mean-square error between the truth and the mean of an ensemble of $M$ independent conditional realisations $\{ \boldsymbol{u}_{u}^{c,m} \}_{m=1}^M$  with $\boldsymbol{u}_{u}^{c,m} \sim \pi(\boldsymbol{u}_u | \boldsymbol{u}_o)$ equals 
\begin{equation}
       \left\langle  \left\|  \frac{1}{M} \sum_{m=1}^M \boldsymbol{u}_u^{c,m} - \boldsymbol{u}_u^{\mathrm{true}} \right\|^2 \right\rangle = \left( 1 + \frac{1}{M} \right)  \operatorname{Tr} \left( \mathbb{E}_{\boldsymbol{u}_{o}} \left[\operatorname{Cov}(\boldsymbol{u}_u | \boldsymbol{u}_o) \right]\right) ,
\end{equation}
which recovers \eqref{eq:mse_conditional_rec} for $M=1$, and the MSE of the ensemble average for $M \rightarrow \infty$, i.e. $ \operatorname{Tr} \left( \mathbb{E}_{\boldsymbol{u}_{o}} \left[\operatorname{Cov}(\boldsymbol{u}_u | \boldsymbol{u}_o) \right] \right)$. 
Note that the trace of the covariance matrix consists of the conditional variances: $[\operatorname{Cov}(\boldsymbol{u}_u | \boldsymbol{u}_o) ]_{i,i} =\operatorname{Var}({u}_{u,i} | \boldsymbol{u}_o ) $.
Consequently, both the minimal obtainable MSE, obtained by the conditional average, and the MSE of realistic conditional fields in \eqref{eq:mse_conditional_rec} are related to the conditional variance.

\subsection{Approximation of the error with linear stochastic estimation}
The conditional covariance matrix is defined through the conditional average, which is unknown. 
Because we typically only have one realisation of the field in the unobserved region per realisation of the observed region available, we cannot compute $\langle \boldsymbol{u}_{u}  | \boldsymbol{u}_{o} \rangle$ from data. Therefore, we will approximate it with the best linear estimator, as given by linear stochastic estimation \citep{adrian_stochastic_1996}.

The conditional average is approximated up to first order as $ \langle \boldsymbol{u}_u | \boldsymbol{u}_o \rangle \approx \langle \boldsymbol{u}_u \rangle +  \mathsfbi{L} (\boldsymbol{u}_o - \langle \boldsymbol{u}_o \rangle)$. 
 The matrix $\mathsfbi{L} \in \mathbb{R}^{(n-m)\times m}$ is determined by minimising the mean-square discrepancy with the conditional average $\langle \boldsymbol{u}_u | \boldsymbol{u}_o \rangle$, averaged over $\boldsymbol{u}_o$. The first-order necessary condition for optimality then requires 
\begin{subeqnarray}\label{eq:lse-fonc}
       0 &=& \frac{\partial}{\partial \mathsfbi{L}}  \langle \| \langle \boldsymbol{u}_u | \boldsymbol{u}_o \rangle - \langle \boldsymbol{u}_u \rangle - \mathsfbi{L} \boldsymbol{u}_o' \|^2 \rangle \\
       &=&  -2 \langle ( \langle \boldsymbol{u}_u | \boldsymbol{u}_o \rangle - \langle \boldsymbol{u}_u \rangle -  \mathsfbi{L}\boldsymbol{u}_o')\boldsymbol{u}_o'^\top  \rangle \\
        &=& -  2  ( \mathsfbi{C}_{uo}-  \mathsfbi{L} \mathsfbi{C}_{oo}  )
\end{subeqnarray}
where (c) follows from the law of total expectation, and introducing $\mathsfbi{C}_{oo}= \langle \boldsymbol{u}_o' \boldsymbol{u}_o'^\top \rangle$ and $\mathsfbi{C}_{uo}= \langle \boldsymbol{u}_u' \boldsymbol{u}_o'^\top \rangle$.
The covariance matrix $\mathsfbi{C}_{oo} $ is positive semidefinite and thus not necessarily invertible. 
In particular, if a discrete incompressibility constraint holds on the velocity field, that is $\mathsfbi{D}\boldsymbol{u}=\boldsymbol{0}$ with $\mathsfbi{D} \in \mathbb{R}^{v\times n}$, then the range of $\mathsfbi{C}$ is constrained to the null space of $\mathsfbi{D}$, so that $\mathsfbi{C}$ is not necessarily of full rank.
Therefore we select $\mathsfbi{L}=\mathsfbi{C}_{uo} \mathsfbi{C}_{oo}^+$, where $\mathsfbi{C}_{oo}^+$ is the Moore--Penrose inverse of $\mathsfbi{C}_{oo}$, thereby making $\mathsfbi{L}$ the minimum norm solution to (\ref{eq:lse-fonc}c).
The best linear estimator for the conditional average is then
\begin{equation}\label{eq:lse}
   \langle \boldsymbol{u}_u | \boldsymbol{u}_o \rangle \approx \langle \boldsymbol{u}_{u} \rangle + \mathsfbi{C}_{uo} \mathsfbi{C}_{oo}^+ (\boldsymbol{u}_{o} - \langle \boldsymbol{u}_{o} \rangle ) .
\end{equation}
Note that the Moore--Penrose inverse of the symmetric matrix $\mathsfbi{C}_{oo}$ projects $\boldsymbol{u}_o' $ onto $\operatorname{col}(\mathsfbi{C}_{oo})$, applies the inverse of $\mathsfbi{C}_{oo}$ restricted to $\operatorname{col}(\mathsfbi{C}_{oo}) $, and embeds the result back into $\mathbb{R}^m$. Since for any $\boldsymbol{v}_o \perp \operatorname{col}(\mathsfbi{C}_{oo})$ holds that $ 0 =\boldsymbol{v}_o^\top \mathsfbi{C}_{oo} \boldsymbol{v}_o = \langle ( \boldsymbol{u}_o'^\top \boldsymbol{v}_o)^2 \rangle$, we find that $\boldsymbol{u}_o' \in \operatorname{col}(\mathsfbi{C}_{oo})$ almost surely, in which case the projection part of $\mathsfbi{C}_{oo}^+$ leaves $\boldsymbol{u}_o'$ unchanged.

With the estimated conditional average from linear stochastic estimation, the conditional covariance matrix, averaged over $\boldsymbol{u}_o$, is approximately given by
\begin{subeqnarray}\label{eq:avg-cov}
        \mathbb{E}_{\boldsymbol{u}_o} [\operatorname{Cov}(\boldsymbol{u}_u | \boldsymbol{u}_o) ]
        &=& \mathbb{E}_{\boldsymbol{u}_o , \boldsymbol{u}_u \sim \pi(\cdot | \boldsymbol{u}_o)} [ (\boldsymbol{u}_u - \langle \boldsymbol{u}_u | \boldsymbol{u}_o \rangle ) (\boldsymbol{u}_u - \langle \boldsymbol{u}_u | \boldsymbol{u}_o \rangle)^\top ] \\
        &\approx& \big\langle (\boldsymbol{u}_u'  - \mathsfbi{C}_{uo} \mathsfbi{C}_{oo}^+\boldsymbol{u}_o') (\boldsymbol{u}_u' -  \mathsfbi{C}_{uo} \mathsfbi{C}_{oo}^+\boldsymbol{u}_o')^\top \big\rangle \\
        &=& \big\langle  \boldsymbol{u}_u' \boldsymbol{u}_u'^\top  -   \mathsfbi{C}_{uo} \mathsfbi{C}_{oo}^+\boldsymbol{u}_o' \boldsymbol{u}_u'^\top - \boldsymbol{u}_u' \boldsymbol{u}_o'^\top \mathsfbi{C}_{oo}^+ \mathsfbi{C}_{ou} + \mathsfbi{C}_{uo} \mathsfbi{C}_{oo}^+\boldsymbol{u}_o' \boldsymbol{u}_o'^\top \mathsfbi{C}_{oo}^+\mathsfbi{C}_{ou} \big\rangle  \\
        &=&  \mathsfbi{C}_{uu}  - \mathsfbi{C}_{uo} \mathsfbi{C}_{oo}^+ \mathsfbi{C}_{ou}
\end{subeqnarray}
where (a) uses the definition of the conditional covariance $\operatorname{Cov}(\boldsymbol{u}_u | \boldsymbol{u}_o)$, (b) approximates the conditional mean with linear stochastic estimation, (c) follows from expanding the squares, and (d) from ensemble averaging over the natural distribution $\boldsymbol{u}\sim \pi (\cdot) $.
Consequently, the total MSE is approximately given by 
\begin{equation}\label{eq:mse-approx}
      \left\langle  \left\|  \frac{1}{M} \sum_{m=1}^M \boldsymbol{u}_u^{c,m} - \boldsymbol{u}_u^{\mathrm{true}} \right\|^2 \right\rangle 
      \approx \left( 1 + \frac{1}{M} \right)   \operatorname{Tr} \left[ \mathsfbi{C}_{uu}  - \mathsfbi{C}_{uo} \mathsfbi{C}_{oo}^+ \mathsfbi{C}_{ou} \right]
      := \left( 1 + \frac{1}{M} \right) \operatorname{Tr} \left[ \mathsfbi{C}_{uu|o}  \right],
\end{equation}
where $\mathsfbi{C}_{uu|o} $ can be recognised as the generalised Schur complement of the full covariance matrix $\mathsfbi{C} = \langle \boldsymbol{u}'\boldsymbol{u}'^\top \rangle$. 
Note that the projection step inherent in the pseudo-inverse of $\mathsfbi{C}_{oo}$ also leaves $\mathsfbi{C}_{ou}$ unchanged, since $\operatorname{col}(\mathsfbi{C}_{ou}) \subset \operatorname{col}(\mathsfbi{C}_{oo}) $. Indeed, $\mathsfbi{C}_{ou} = \langle \boldsymbol{u}_o'\boldsymbol{u}_u'^\top \rangle$, so its columns lie in the span of $ \boldsymbol{u}_o'$, which coincides with $ \operatorname{col}(\mathsfbi{C}_{oo})$.

As in spectral linear stochastic estimation \citep{tinney_spectral_2006}, we make use of the property that the Fourier-transformed covariance matrices diagonalise per wavenumber in the statistically homogeneous directions. This reduces the size of the covariance matrices that must be inverted and improves the numerical robustness when estimating the covariances from data.
Although the channel flow is statistically homogeneous in both the streamwise and spanwise direction, only the spanwise homogeneity is preserved by the observational setup shown in figure~\ref{fig:reconstruction}(a). 
Therefore, we will only consider the spanwise Fourier transform.

To show that the conditional covariance matrix diagonalises per spanwise wavenumber, let the state vector be ordered by spanwise planes: $\boldsymbol{u}= \{\boldsymbol{u}^{(i_z)}\}_{i_z=0}^{n_z-1}$, where $\boldsymbol{u}^{(i_z)} \in \mathbb{R}^{q}$ and $q=3 n_x n_y$. 
With this convention, the unitary discrete Fourier transform in the spanwise direction is $\mathsfbi{F} =  \mathsfbi{F}_z  \otimes \mathsfbi{I}_q$ with $\mathsfbi{F}_z \mathsfbi{F}_z^\ast = \mathsfbi{I}_{n_z}$, where $\otimes$ denotes the Kronecker product and the asterisk the conjugate transpose. 
Hence, the spanwise Fourier representation of the covariance matrix $\mathsfbi{C}= \langle \boldsymbol{u}'\boldsymbol{u}'^\top \rangle $  is $\mathsfbi{\hat{C}} = \mathsfbi{F} \mathsfbi{C} \mathsfbi{F}^\ast$. 
Under the assumptions of periodicity and statistical homogeneity in the spanwise direction, $\mathsfbi{\hat{C}}$ is block diagonal and can be written as the direct sum 
\begin{equation}
    \mathsfbi{\hat{C}} = \operatorname{diag}(  \mathsfbi{\hat{C}}^{(0)},   \mathsfbi{\hat{C}}^{(1)},   \ldots, \mathsfbi{\hat{C}}^{(n_z-1)}) \equiv \bigoplus_{k_z=0}^{n_z-1} \mathsfbi{\hat{C}}^{(k_z)},
\end{equation}
where $\mathsfbi{\hat{C}}^{(k_z)} \in \mathbb{C}^{q\times q}$ is the covariance matrix associated with wavenumber-index $ k_z$.
For the measurement setup shown in figure~\ref{fig:reconstruction}(a), the mask matrix can be written as $\mathsfbi{M}= \operatorname{diag}(\mathsfbi{M}_r, \ldots, \mathsfbi{M}_r) \equiv \mathsfbi{I}_{n_z} \otimes \mathsfbi{M}_r$ where $\mathsfbi{M}_r \in \mathbb{R}^{r \times q}$ with $r = m/n_z$. The complementary mask matrix is then given by $\mathsfbi{M}^\perp =\mathsfbi{I}_{n_z} \otimes \mathsfbi{M}_r^\perp $ where $\mathsfbi{M}_r^\perp \in \mathbb{R}^{p \times q}$ and $p=q-r$. 
Because the mask is uniform along the spanwise direction, it follows that $\mathsfbi{M} \mathsfbi{F} = \mathsfbi{F}_o \mathsfbi{M}$ where $ \mathsfbi{F}_o = \mathsfbi{F}_z \otimes  \mathsfbi{I}_r  $ and $\mathsfbi{M}^\perp \mathsfbi{F} = \mathsfbi{F}_u \mathsfbi{M}^\perp$ where $ \mathsfbi{F}_u =  \mathsfbi{F}_z \otimes \mathsfbi{I}_{p} $. 
Consequently, the Fourier-transformed covariance submatrices inherit the same block-diagonal structure. For example, we find for $\mathsfbi{C}_{uo}=\mathsfbi{C}_{ou}^\ast$ that
\begin{equation*}
     \mathsfbi{{\hat{C}}}_{uo} = \mathsfbi{F}_u \mathsfbi{M}^\perp \mathsfbi{C} \mathsfbi{M}^\top \mathsfbi{F}_o^\ast
    = \mathsfbi{M}^\perp \mathsfbi{F} \mathsfbi{C} \mathsfbi{F}^\ast \mathsfbi{M}^\top  
    = \left(  \mathsfbi{I}_{n_z} \otimes \mathsfbi{M}_r^\perp \right) \left( \bigoplus_{k_z=0}^{n_z-1} \mathsfbi{\hat{C}}^{(k_z)} \right) \left(  \mathsfbi{I}_{n_z} \otimes \mathsfbi{M}_r^\top \right)     = \bigoplus_{k_z=0}^{n_z-1} \mathsfbi{\hat{C}}_{uo}^{(k_z)} .
\end{equation*}
As a result, the Fourier-transformed conditional covariance, $\mathsfbi{\hat{C}}_{uu|o} =\mathsfbi{F}_u \mathsfbi{C}_{uu|o} \mathsfbi{F}_u^\ast$, becomes
\begin{equation}\label{eq:cond-var-gauss}
        \mathsfbi{\hat{C}}_{uu|o}
    =  \bigoplus_{k_z=0}^{n_z-1} \left[ \mathsfbi{\hat{C}}_{uu}^{(k_z)} -  \mathsfbi{\hat{C}}_{uo}^{(k_z)}  \left( \mathsfbi{\hat{C}}_{oo}^{(k_z)} \right)^+  \mathsfbi{\hat{C}}_{ou}^{(k_z)} \right]
     := \bigoplus_{k_z=0}^{n_z-1}  \mathsfbi{\hat{C}}_{uu|o}^{(k_z)}.
\end{equation}

The mean-square reconstruction error can be computed directly from the blocks of the Fourier-transformed conditional covariance matrix.  
Since the spanwise homogeneity also holds for the reconstruction error, we directly compute the spanwise average with Parseval's theorem.
Let $u_{u,\ell}^{(i_z)}$ denote the $\ell$-th unobserved velocity component in the $i_z$-th spanwise plane, where $\ell=0,\ldots,p-1$. Equivalently, the corresponding global index is $i=i_zp+\ell$. The spanwise-averaged mean-square reconstruction error for this component is
\begin{equation}
    \varepsilon_\ell^2  
    := \frac{1}{n_z} \sum_{i_z=0}^{n_z-1} \left\langle \left( \frac{1}{M} \sum_{m=1}^{M}  u_{u,\ell}^{c,m,(i_z)} - u_{u,\ell}^{\mathrm{true},(i_z)}  \right)^2  \right\rangle
    \approx  \left( 1+\frac{1}{M} \right) \frac{1}{n_z} \sum_{i_z=0}^{n_z-1}  \left[ \mathsfbi{C}_{uu|o}^{(i_z)} \right]_{\ell,\ell},
\end{equation}
where $\mathsfbi{C}_{uu|o}^{(i_z)}\in\mathbb{R}^{p\times p}$ denotes the block matrix on the diagonal of $\mathsfbi{C}_{uu|o}$ associated with spanwise plane $i_z$. Using Parseval's identity for the unitary spanwise DFT, we then find
\begin{equation}
    \varepsilon_\ell^2
    \approx 
   \left( 1+\frac{1}{M} \right) \frac{1}{n_z} \sum_{k_z=0}^{n_z-1} \left[ \mathsfbi{\hat{C}}_{uu|o}^{(k_z)} \right]_{\ell,\ell}.
\end{equation}
This expression is used to compute the normalised MSE in \eqref{eq:normalised-mse}, shown in figures~\ref{fig:reconstruction}(b-d). 
Similarly, we find that the total MSE in \eqref{eq:mse-approx} can be computed without an inverse Fourier transform as
\begin{equation}
  \left\langle  \left\|  \frac{1}{M} \sum_{m=1}^M \boldsymbol{u}_u^{c,m} - \boldsymbol{u}_u^{\mathrm{true}} \right\|^2 \right\rangle 
      \approx \left( 1 + \frac{1}{M} \right) \sum_{k_z=0}^{n_z-1} \operatorname{Tr} \left( \mathsfbi{\hat{C}}_{uu|o}^{(k_z)}\right),
\end{equation}
which follows from the cyclic property of the trace $\operatorname{Tr}( \mathsfbi{C}_{uu|o} ) = \operatorname{Tr}(\mathsfbi{F}_u  \mathsfbi{F}_u^\ast \mathsfbi{\hat{C}}_{uu|o} ) = \operatorname{Tr}(\mathsfbi{\hat{C}}_{uu|o})$ and the block-diagonal structure of $\mathsfbi{\hat{C}}_{uu|o}$.

\subsection{Estimation of the conditional covariance from DNS snapshots}
For each spanwise wavenumber index $k_z$, we estimate the corresponding block of the Fourier-transformed generalised Schur complement $\mathsfbi{\hat{C}}_{uu|o}^{(k_z)}$ from DNS snapshots. To arrive at an unbiased estimator for $\mathsfbi{\hat{C}}_{uu|o}^{(k_z)}$, note that the linear stochastic estimator in \eqref{eq:lse} can be interpreted as the conditional mean of a Gaussian approximation to the natural distribution of the velocity field $\boldsymbol{u} \sim \mathcal{N}(\langle \boldsymbol{u}\rangle , \mathsfbi{C})$. For a possibly singular positive-semidefinite covariance matrix, the conditional Gaussian distribution is defined as (\citealt{ouellette_schur_1981}, §6.1)
\begin{equation}
    \boldsymbol{u}_u |  \boldsymbol{u}_o  \sim \mathcal{N} \left(\langle \boldsymbol{u}_{u} \rangle + \mathsfbi{C}_{uo} \mathsfbi{C}_{oo}^+ (\boldsymbol{u}_{o} - \langle \boldsymbol{u}_{o} \rangle ) , \mathsfbi{C}_{uu}  - \mathsfbi{C}_{uo} \mathsfbi{C}_{oo}^+ \mathsfbi{C}_{ou} \right),
\end{equation}
where we use the Moore--Penrose inverse as generalised inverse. 
Under this Gaussian assumption, the non-self-conjugate spanwise Fourier coefficients follow the complex conditional Gaussian distribution  
\begin{equation}\label{eq:complex-gauss}
    \boldsymbol{\hat{u}}_u^{(k_z)} |  \boldsymbol{\hat{u}}_o^{(k_z)}  \sim \mathcal{CN} \left(\langle  \boldsymbol{\hat{u}}_u^{(k_z)} \rangle + \mathsfbi{\hat{C}}_{uo}^{(k_z)} \mathsfbi{\hat{C}}_{oo}^{(k_z)^+} \boldsymbol{\hat{u}}_{o}'^{(k_z)} , \mathsfbi{\hat{C}}_{uu}^{(k_z)}  - \mathsfbi{\hat{C}}_{uo}^{(k_z)} \mathsfbi{\hat{C}}_{oo}^{(k_z)^+}  \mathsfbi{\hat{C}}_{ou}^{(k_z)} \right),
\end{equation}
whereas the self-conjugate coefficients follow the corresponding real conditional Gaussian distribution. 
Consequently, we can use an unbiased estimator for the conditional covariance in \eqref{eq:complex-gauss} to estimate $\mathsfbi{\hat{C}}_{uu|o}^{(k_z)}$.

Consider $N_s$ independent DNS snapshots $\{ \boldsymbol{u}_s\}_{s=1}^{N_s}$. The corresponding centred snapshots are defined by $\{ \boldsymbol{u}_s'\}_{s=1}^{N_s}$ where $\boldsymbol{u}_s' = \boldsymbol{u}_s - \sum_{s=1}^{N_s} \boldsymbol{u}_s/N_s $.
Let $ \boldsymbol{\hat{u}}_{o,s}'^{(k_z)} \in \mathbb{C}^{r}$ denote the spanwise discrete Fourier transform of $\mathsfbi{M}\boldsymbol{u}_s'$ at wavenumber-index $k_z$, and let $ \boldsymbol{\hat{u}}_{u,s}'^{(k_z)}\in \mathbb{C}^{p}$ be defined analogously for $\mathsfbi{M}^\perp \boldsymbol{u}_s'$.
These vectors are collected in the snapshot matrices 
\[
\mathsfbi{\hat{U}}_{o}'^{(k_z)}
=
\left(
\boldsymbol{\hat{u}}_{o,1}'^{(k_z)},\ldots,
\boldsymbol{\hat{u}}_{o,N_s}'^{(k_z)}
\right)
\in\mathbb{C}^{r\times N_s},
\qquad
\mathsfbi{\hat{U}}_{u}'^{(k_z)}
=
\left(
\boldsymbol{\hat{u}}_{u,1}'^{(k_z)},\ldots,
\boldsymbol{\hat{u}}_{u,N_s}'^{(k_z)}
\right) \in \mathbb{C}^{p \times N_s}.
\]
The resulting unbiased sample covariance estimators for $\mathsfbi{\hat{C}}_{uu}^{(k_z)}$,  $\mathsfbi{\hat{C}}_{oo}^{(k_z)}$, and $\mathsfbi{\hat{C}}_{uo}^{(k_z)}$ are 
\[
\mathsfbi{\hat{S}}_{uu}^{(k_z)}= \frac{1}{N_s-1} \mathsfbi{\hat{U}}_{u}'^{(k_z)} \mathsfbi{\hat{U}}_{u}'^{(k_z)\ast} , \qquad
\mathsfbi{\hat{S}}_{oo}^{(k_z)}= \frac{1}{N_s-1}  \mathsfbi{\hat{U}}_{o}'^{(k_z)} \mathsfbi{\hat{U}}_{o}'^{(k_z)\ast} , \qquad
\mathsfbi{\hat{S}}_{uo}^{(k_z)}= \frac{1}{N_s-1} \mathsfbi{\hat{U}}_{u}'^{(k_z)} \mathsfbi{\hat{U}}_{o}'^{(k_z)\ast} .
\]
Under the Gaussian approximation and with $\rho_{k_z} =\operatorname{rank}(\mathsfbi{\hat{S}}_{oo}^{(k_z)})$, the generalised Schur complement estimator 
\begin{equation}
    \mathsfbi{\hat{S}}_{uu|o}^{(k_z)}
    =
    \frac{N_s-1}{N_s-1-\rho_{k_z}}
    \left[
    \mathsfbi{\hat{S}}_{uu}^{(k_z)}
    -
    \mathsfbi{\hat{S}}_{uo}^{(k_z)}
    \left(
    \mathsfbi{\hat{S}}_{oo}^{(k_z)}
    \right)^+
    \mathsfbi{\hat{S}}_{ou}^{(k_z)}
    \right],
    \label{eq:estimator-schur}
\end{equation}
can be shown to be unbiased, i.e. $ \mathbb{E} [  \mathsfbi{\hat{S}}_{uu|o}^{(k_z)}  ] = \mathsfbi{\hat{C}}_{uu|o}^{(k_z)} $, based on Theorem 3.3 of \cite{andersen_linear_1995}. 
The finite-sample correction factor is consistent with the rank-dependent correction of \cite{ouellette_schur_1981} §6.4 in the real case. 
When $\mathsfbi{\hat{S}}_{oo}^{(k_z)}$ is full rank, the complex positive-definite result of \cite{andersen_linear_1995} Theorem 3.6 is recovered.
To keep $ \rho_{k_z} \leq r$ well below $N_s-1$, we downsample the velocity field by an additional factor 4 in the wall-normal direction, when estimating the generalised Schur complement. Because of the stretched grid in this direction, the first- and second-order statistics are found to be still adequately represented after downsampling.

\end{appen}\clearpage

\bibliographystyle{bibstyle}
\bibliography{bib}

\end{document}